\documentclass[onecolumn,authoryear]{els-mrw} 

\usepackage{amsmath,amssymb,amsfonts,amsthm,makeidx,graphicx}
\usepackage{txfonts}
\usepackage{helvet}

\begin{document}

\chapter{Liquid-gas phase transition of nuclear matter}\label{chap1}

\author[1,2]{Norbert Kaiser}
\author[1,2]{Wolfram Weise}
\address[1]{\orgname{Technical University of Munich}, \orgdiv{School of Natural Sciences, Physics Department}, \orgaddress{85748 Garching, Germany}}
\address[2]{\orgdiv{Excellence Cluster ORIGINS}, \orgaddress{Boltzmannstr. 2, 85748 Garching, Germany}}

\articletag{Chapter Article tagline ..}
\maketitle

\begin{glossary}[Nomenclature]
\begin{tabular}{@{}lp{34pc}@{}}
ChEFT & Chiral Effective Field Theory\\
ChPT & Chiral PerturbationTheory\\
EoS & Equation of State\\
FRG & Functional Renormalization Group\\
HF & Hartree-Fock\\
LGT & Liquid-GasTransition\\
MF & Mean-Field\\
QCD & Quantum Chromodynamics\\
VdW & Van der Waals\\
\end{tabular}
\end{glossary}

\begin{abstract}[Abstract]
A survey is presented summarizing the empirical evidence for and interpretations of a first-order liquid-gas phase transition in nuclear matter.  Earlier developments and the present state of knowledge about the extraction of the critical point for such a transition,  primarily from the systematics of multifragmention data,  are outlined.  By analogy with a Van der Waals equation of state, the empirically deduced critical temperature and pressure permit to draw a schematic picture of the underlying nuclear potential.  More detailed approaches to the liquid-gas transition using self-consistent nuclear Hartree-Fock and variational calculations are described.  Critical exponents are reported. Then chiral effective field theory,  as the low-energy realization of QCD, is discussed in the context of nuclear thermodynamics.  Its implications for the liquid-gas transition in symmetric nuclear matter as well as in neutron-rich matter are reviewed. 
\end{abstract}

\section{Introduction and overview}\label{ch1:Intro}

The liquid-gas transition is a prototypic example of a phase transition in statistical physics.  It is known to exist for fluids with two-body interactions characterized by a combination of short-range repulsion and long-range attraction.  Non-ideal gases of atoms and molecules have such features,  as described for example quite sucessfully by a Van der Waals equation of state.  Nuclear two-body potentials are known to have qualitatively similar properties at the corresponding femtometer (fm) scale: a strongly repulsive core at short distance between nucleons,  accompanied by an attractive interaction at larger distances.  The quest for a transition from liquid to vapor in bulk nuclear matter has thus been a key topic in nuclear physics ever since (Boal, 1986; Csernai {\it et al.}, 1986; Durand {\it et al.}, 2001,  and refs.  therein) 

Nuclear matter is an idealised, infinite many-body system of interacting protons and neutrons.  It represents the bulk part of nuclei, ignoring the nuclear surface and Coulomb effects.  A first-order phase transition in the strictest sense can exist only in an infinite system,  although the notion of phase transitions in systems with a finite number of constituents can also be defined and is as such well established (e.g. Chomaz {\it et al.}, 2002).  Nuclei are finite objects with proton number $Z$ and neutron number $A-Z$ (total baryon number $A$),  localized in a finite volume $V$.  It is indeed possible to relate the existence of a liquid-gas phase transition in nuclear matter to an instability in excited finite nuclei with respect to the evaporation of nucleons and the formation of fragments (clusters). Such multifragmentation processes are empirically verified in intermediate-energy nuclear collisions. Concepts of temperature and entropy (nuclear thermodynamics) can be introduced at low excitation energies (Gross, 1990; Borderie {\it et al.}, 2019; Bonnet {\it et al.}, 2025).  The focus is on energies per nucleon in the range $E/A\lesssim 30$ MeV, corresponding to temperatures $T\lesssim 25$ MeV,  together with bulk nuclear densities, $n = A/V$,  below the equilibrium density of saturated nuclear matter,  $n_0 \simeq 0.16$ fm$^{-3}$, at $T=0$.

By systematically sampling charged fragments from nuclear collisions in this energy range and correcting for surface and Coulomb effects, it is possible to extract a critical temperature, $T_c$,  that marks the end point of a liquid-vapor coexistence region\footnote{The term liquid-vapor transition is sometimes used in this context as a reminder that there is not just a gas of nucleons but also a variety of composite nuclear fragments in the final state.  We shall nevertheless generally stay with the term liquid-gas transition (LGT) as refering to all possible evaporated fragments.} (Elliot {\it et al.}, 2013).  Signatures of such a coexistence region are found in measured caloric curves,  plotting temperature as a function of excitation energy (Borderie {\it et al.}, 2008, 2023).  
A survey of results and evidence for a nuclear first-order liquid-gas phase transition follows in Chapter \ref{ch2}.

It is instructive to compare the empirically deduced parameters for the critical point of the nuclear liquid-gas transition (LGT)  with a Van der Waals equation of state by analogy.  Indeed,  the long-range attractive interaction that governs the nuclear LGT has a dominant two-pion exchange part which is reminiscent of a Van der Waals (VdW) potential,  the difference being that the characteristic $1/r^6$ leading behaviour of the atomic VdW interaction is now modified by an exponential screening factor involving the pion mass.  By examining the second virial coefficient for the pressure a qualitative picture of the underlying potential can be drawn as will be shown in Chapter \ref{ch3}.

Calculations of the nuclear matter equation of state at finite temperatures with reference to the liquid-gas transition (LGT) have a long history,  often inspired by the aim to understand the dynamics of supernova explosions (Lamb {\it et al.}, 1978; Siemens, 1983; Panagiotou {\it et al.}, 1984).  Descriptions of the nuclear LGT frequently employed phenomenological energy density functionals.  Representative examples are given in Chapter \ref{ch4}.  A contemporary theoretical framework with closer links to QCD is Chiral Effective Field Theory (ChEFT).  It is based on the spontaneously broken chiral symmetry of QCD at low energies.  Pions emerge as (approximate) Goldstone bosons in this theory.  ChEFT emphasizes the special role of pions in deriving nuclear interactions and nuclear thermodynamics (Wellenhofer {\it et al.}, 2014, 2015; Holt {\it et al.}, 2016; Drischler {\it et al.},  2021) as outlined in Chapter \ref{ch5}.  Some aspects of the LGT viewed from the broader context of the phase diagram for strongly interacting matter are discussed in Chapter \ref{ch6},  followed by a concluding summary in Chapter \ref{summary}.

(Note: Throughout this article,  temperatures will be expressed in units of the Boltzmann constant (i.e.  $k_B = 1$).  We also work with the commonly used action and velocity units in nuclear and particle physics,  i.e. $\hbar = c = 1$.)

\section{The quest for a nuclear liquid-gas phase transition}\label{ch2}

Nuclei containing sufficiently large numbers of protons and neutrons can be viewed as droplets of a Fermi liquid.  Collisions of such nuclei at low and intermediate energies create excited states characterized by their excitation energy per nucleon, $E^*/A$.  The system is heated up,  leading to the evaporation of nucleons and clusters of lighter nuclei.  The observed systematics of these fragments detected in the final states of nuclear collisions (multifragmentation) provides information and signals that suggest the existence of a phase transition in bulk nuclear matter.

\subsection{The data base: multifragmentation}

Multifragmentation refers to the copious emission of fragments and particles when the nuclear excitation energy becomes comparable to the binding energy (Bertsch {\it et al.}, 1983),  i.e.  for excitation energies per nucleon in the range $E^*/A \sim 3-10$ MeV.  At higher excitation energies vaporization of nuclei takes over.  Relating the observed yields of fragments to thermodynamic concepts under the assumption of thermal and chemical equilibrium requires to consider a number of non-trivial corrections accounting for the finiteness of the nuclear systems involved,  in particular surface and Coulomb effects,  and the detailed reaction dynamics of the colliding nuclei (Bondorf {\it et al.}, 1985; Borderie {\it et al.},  2008).  For example,  the critical temperature of an anticipated first-order phase transition in bulk nuclear matter is expected to be significantly higher than that of a LGT in finite systems, since surface effects and long-range Coulomb interactions between protons lower the nuclear binding energy per particle from about 16 MeV in symmetric nuclear matter to roughly half of that value.

\begin{figure}[t]
\centering
\includegraphics[width=.5\textwidth]{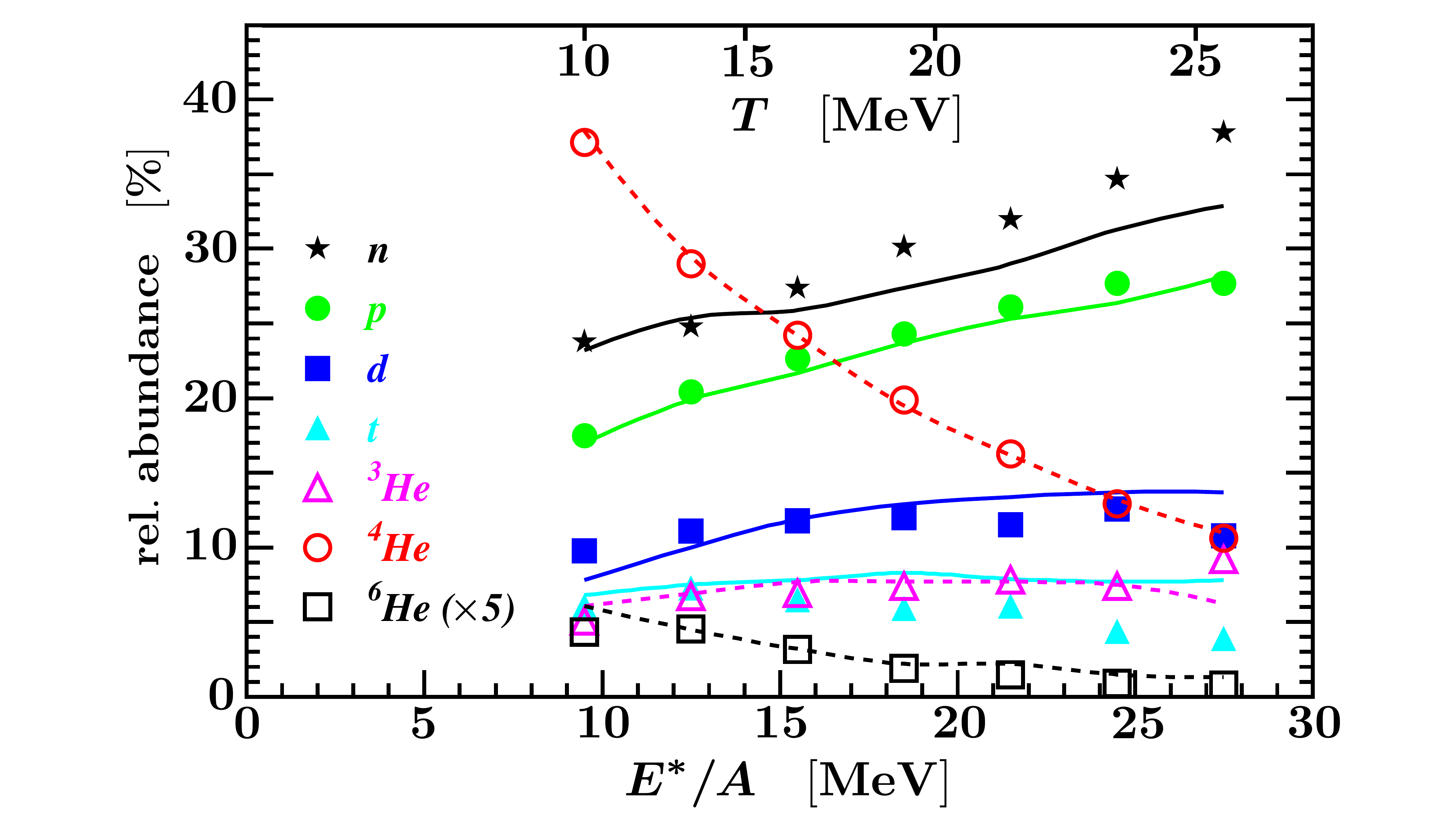}
\caption{Relative abundances of neutrons,  $H$ and $He$ isotopes as a function of the excitation energy per nucleon of vaporized sources produced in collisions of $^{36}Ar$ and $^{58}Ni$ nuclei.  Data are taken with the INDRA multidetector at GANIL.  Curves represent fits using statistical (thermal) distributions of the different particle species after freeze-out.  The corresponding (upper) temperature scale is also shown. (Adapted from Borderie {\it et al.},  2002).  }
\label{Fig1}
\end{figure}

The validity of equlibrium assumptions for vaporizing sources was tested in representative studies,  e.g.  at GANIL using the INDRA multidetector (B. Borderie {\it et al.}, 1999, 2008).  The deexitation into light fragments following 
the production of excited sources in nuclear collisions was analysed starting from a quantum statistical model.  It was demonstrated that,  even for comparatively small numbers of particles,  standard rules of equilibrium thermodynamics are applicable.  An example is given in Fig.\,\ref{Fig1}.  It shows how the relative abundances of light fragments emerging after freeze-out from two colliding nuclei can be successfully reproduced using quantum statistical methods.  It also shows how the excitation energy per nucleon in the vaporizing system translates into a temperature scale.

\begin{figure}[t]
\centering
\includegraphics[width=.7\textwidth]{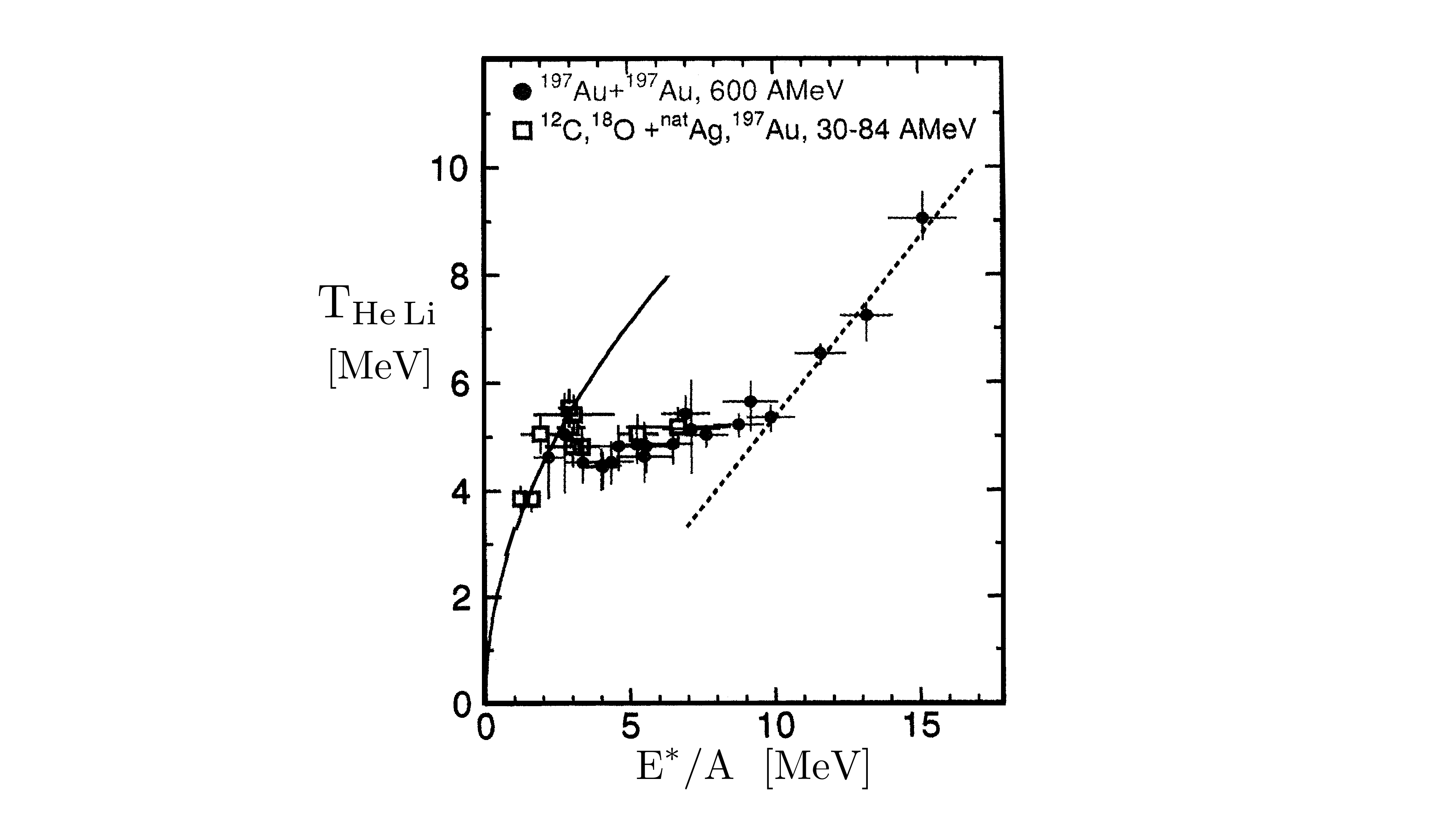}
\caption{Caloric curve (temperature vs. excitation energy per nucleon) determined from different multifragmentation reactions as indicated.  The temperature scale $T=T_{He\,Li}$ is derived from yield ratios of $He$ and $Li$ isotopes.  Fit curves: $T=\sqrt{k\,E^*/A}$ with $k=10$ MeV corresponding to liquid phase (solid curve); $T={2\over 3}(E^*/A -\delta)$ with $\delta = 2$ MeV corresponding to gas phase (dashed curve). Figure adapted from Pochodzalla {\it et al.} (1995).}
\label{Fig2}
\end{figure}

{\bf Excitation energy and temperature: the caloric curve.} The relation between excitation energy and temperature is at the basis of establishing a connection between multifragmentation measurements and a LGT signal.  Consider a collision event leading to intermediate sources of vaporization with binding energies $B_s$.  These sources decay into final fragments of different nuclear species with kinetic energies $K_i$ and binding energies $B_i$.  The excitation energy $E^*$ assigned to the vaporizing source is given by:
\begin{eqnarray}
E^* = \sum_i \left(K_i + B_i + \delta V_i^{\text{coul}}\right) - B_s ~,
\end{eqnarray}
where the sum goes over all detected fragments.  The correction term $ \delta V_i^{\text{coul}}$ takes into account effects of the Coulomb interaction among charged particles.  A method to set the temperature scale is by taking double ratios of isotope pairs differing by one neutron each.  An example of such a setup was realized in the ALADIN measurements at GSI (Pochodzalla {\it et al.}, 1995) where a temperature
scale denoted $T_{He\,Li}$ was introduced proportional to the inverse logarithm of a ratio of yield ratios, $[Y(^6Li)/ Y(^7Li)]/[Y(^3He)/Y(^4He)]$.  A microcanonical multifragmentation model (Gross, 1993) reproduces (to within about 15\%) the temperature scale determined in this way.
At low temperatures,  $T\lesssim 3$ MeV,  and low densities one expects a behaviour of temperature vs.  excitation energy as 
\begin{eqnarray}
T \simeq \sqrt{k{E^*\over A}}~,
\end{eqnarray}
characteristic of a liquid,  where $k$ is a parameter related to the inverse level density.  At higher temperatures beyond the onset of vaporization,  a linear relation typical of a gas,
\begin{eqnarray}
T \simeq {2E^*\over 3A}~,
\end{eqnarray}
is expected.  Such a behaviour is observed in the empirical caloric curve, $T(E^*)$, shown in Fig.\,\ref{Fig2}.  It suggests a transition from a low-temperature (liquid) phase followed by a ``freeze-out" region of almost constant temperature to a phase at higher temperature which is consistent with the linear behaviour, $T\propto E^*$,  of a gas. The intermediate range of constant temperature is interpreted as the region where liquid and gas coexist.

Over the years much progress has been made collecting data of caloric curves in a variety of systems,  strengthening the evidence for a liquid-gas coexistence region and the nuclear LGT (Borderie {\it et al.}, 2013; Bonnet {\it et al.},  2025).  Analysing the observed systematics it has become possible to locate a critical point and its parameters,  and to develop further insights into the nuclear many-body dynamics behind the LGT.

\subsection{Extraction of the critical point}
Studies of evaporation rates from a fluid system such as water inform about its saturated vapour in equilibrium. Thermodynamics determines a phase diagram via measurements of temperature $T$ and pressure $P$ at which the fluid and gas phases coexist.  In nuclear systems the situation is far more complex.  The clusters emitted from the excited nuclei carry the information about the liquid and its transition to vapor.  But the steps to translate observed nuclear evaporation rates into conclusions about the LGT in an infinite system must rely on sophisticated methods to correct for surface and Coulomb effects.  

Combinations of a variety of different data sets from compound nuclear reactions and multifragmentation processes have been analysed (Elliot {\it et al.}, 2013),  accounting for finite-size and Coulomb effects using a liquid-drop ansatz that permits a separation of bulk volume and surface parts.  Corrections for angular momentum arising in the nuclear collisions are also dealt with.  Systematic fits to yields of charged fragments from all those reactions as a function of excitation energy can then be combined into a single scaled curve describing the behaviour of bulk nuclear matter.  These fits permit to obtain an estimate for the critical temperature, $T_c$,  of the LGT in the infinite system.  Assuming ideal gas conditions for the asymptotically detected clusters,  values for the critical density, $n_c$, and critical pressure, $P_c$, can also be extracted.  The empirical critical end point of the liquid-gas coexistence region in the phase diagram of symmetric nuclear matter is then deduced to be located at  (Elliot {\it et al.}, 2013):
\begin{eqnarray}
T_c =  17.9\pm 0.4 ~\text{MeV}~,~~~~P_c = 0.31\pm 0.07~\text{MeV/fm}^3~,~~~~ n_c = 0.06\pm 0.01~\text{fm}^{-3}~.
\label{eq:critpoint}
\end{eqnarray}
(An updated,  larger uncertainty of the critical density has been reported by Moretto et al.  (2017): $n_c = 0.06\pm 0.02~\text{fm}^{-3}$).

\section{Analogies with a Van der Waals equation of state}\label{ch3}

It is instructive to compare the empirically deduced critical parameters (\ref{eq:critpoint}) with those of a Van der Waals (VdW) equation of state.  We recall that this EoS was designed to describe non-ideal atomic or molecular gases and their first-order liquid-vapor transitions.  The underlying atomic/molecular potential is a combination of strong short-range repulsion and an attractive longer-range interaction,  often represented as a Lenard-Jones potential in the form ${\cal V}(r) = V_0[(R/r)^{12}-(R/r)^6]$ with parameters $V_0$ and $R$ adjusted to the species in question.  The attractive part with its characterisic $r^{-6}$ behaviour (the Van der Waals potential) derives from two-photon exchange processes between the two atoms,  involving an intermediate excited state of the system.

The Van der Waals EoS expresses the pressure at temperature $T$ and density $n = N/V$ (where $N$ is the fixed number of constituent particles and $V$ the volume) as follows:
\begin{eqnarray}
P(V,T) = {nT\over 1-b\,n}-a\,n^2 = {T\over \varv-b}-{a\over \varv^2}~,   ~~~~(\varv=1/n= V/N)~.
\label{eq:vdw}
\end{eqnarray}
The parameters $a$ and $b$ reflect the long-range attractive and short-range repulsive interactions,  respectively,  and their effects as they modify the ideal gas EoS.  The latter is recovered for $a=b=0$.  The term $-a\,n^2$ reduces the pressure proportional to the square of the density as expected for an attractive two-body interaction.  The parameter $b$ suggests an interpretation as a rough measure of the excluded volume per particle that cannot be occupied by the constituents because their finite sizes and strongly repulsive interactions at short distance keep them mutually apart.  

{\bf Critical point.} The VdW EoS features a first-order liquid-gas phase transition after a Maxwell construction,  with a critical point (saddle point) determined by ${\partial P\over\partial V} ={\partial^2 P\over\partial V^2} = 0$ at $T = T_c$.  The resulting conditions,
\begin{eqnarray}
-{T_c\over (\varv_c - b)^2} + {2a\over \varv_c^3} = 0~, ~~~\text{and} ~~~{T_c\over (\varv_c - b)^3} + {3a\over \varv_c^4} = 0~,
\end{eqnarray}
lead to relations between $a$, $b$ and the values of temperature, pressure and density at the critical pont:
\begin{eqnarray}
T_c = {8\,a\over 27\,b}~,~~~~~P_c = {a\over 27\,b^2}~,~~~~~n_c={1\over \varv_c}={1\over 3\,b}
\end{eqnarray}
The following dimensionless ratio is characteristic of the VdW EoS:
\begin{eqnarray}
{T_c\,n_c\over P_c }= {8\over 3}~,~~~~~\text{or}~~~~~{P_c\over T_c} = 0.375\,n_c~.
\end{eqnarray}

{\bf Specific heat.} The internal energy $E$ of a VdW gas is taken as independent of $b$.  It differs from the internal energy $E_{id} = {3\over 2}NT$ of an ideal gas only by the attractive term proportional to $a$ in the EoS (\ref{eq:vdw}):
\begin{eqnarray}
{E\over V} = {E_{id}\over V} - n^2\,a~.
\end{eqnarray}
Consequently, the specific heat $C_V = (\partial E/\partial T)_V$ at constant volume coincides with that of an ideal gas.  At constant pressure $P$,  the absorbed thermal energy is related to the enthalpy,  $H = E + PV$, and the corresponding specific heat is $C_P = (\partial H/\partial T)_P$. The difference $C_P - C_V$ is a quantity of interest when examining the thermodynamics of the non-ideal system.  For the Van der Waals EoS one finds:
\begin{eqnarray}
C_P - C_V = {N\over 1-2n a(1-nb)^2/T}~.
\end{eqnarray}
When $T \rightarrow T_c$ and $n \rightarrow n_c$,  inserting the values $a, b$ one confirms that $C_P\rightarrow\infty$ at the critical point.  

\begin{figure}[t]
\centering
\includegraphics[width=.5\textwidth]{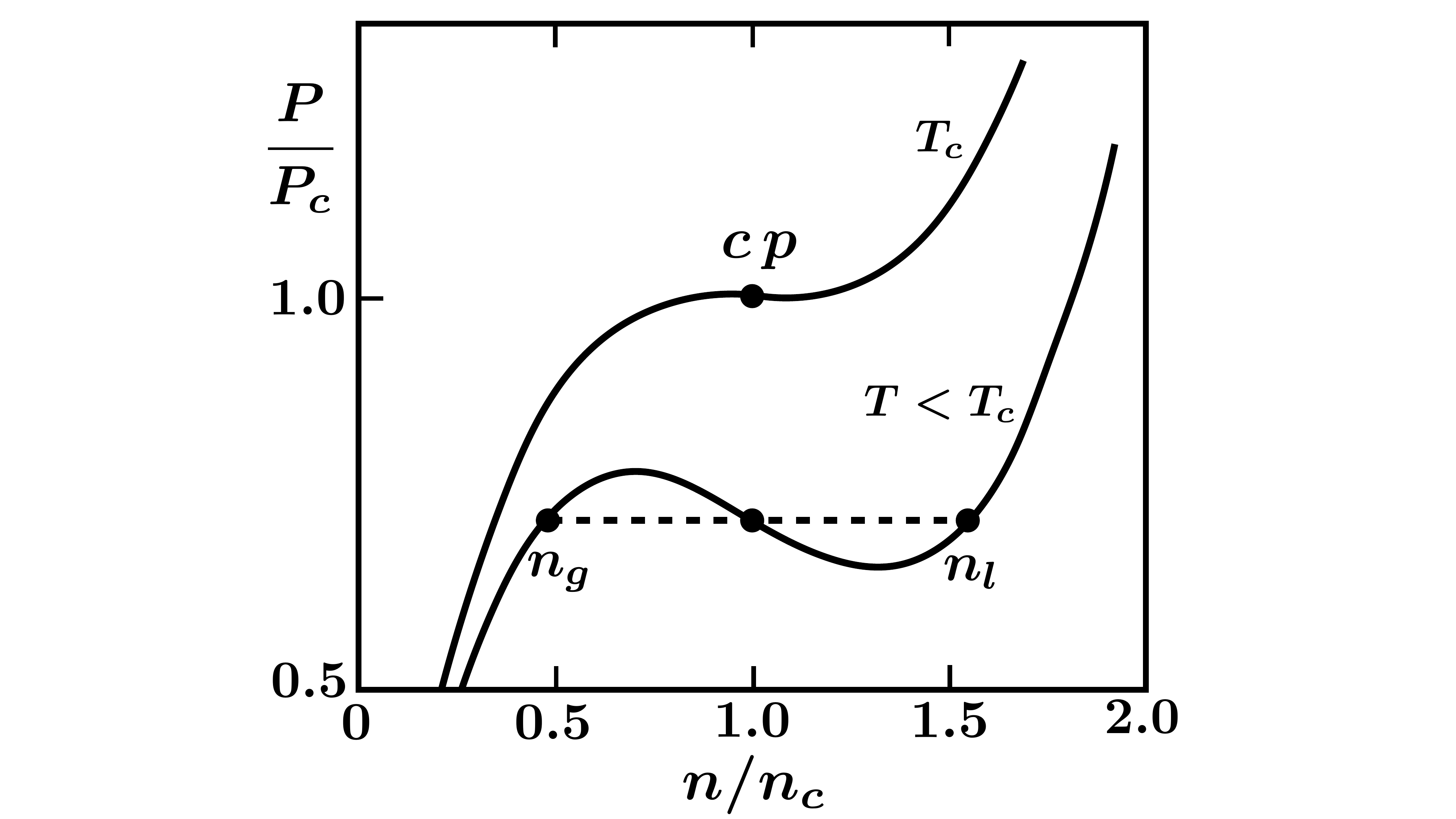}
\caption{Two isotherms of a Van der Waals type equation-of-state $P(n,T)$ in the vicinity of the critical point ($cp$).  The starting and end points of the gas-liquid phase coexistence region for $T<T_c$ are marked by their densities $n_g$ and $n_l$, respectively.  The horizontal dashed line illustrates the Maxwell construction.  }
\label{Fig3}
\end{figure}

{\bf Density fluctuations at the critical point.} The isothermal compressibility,  
\begin{eqnarray}
K_T=-{1\over V}\left({\partial P\over\partial V}\right)_T^{-1} = {V\over T}\left({\Delta N\over\langle N\rangle}\right)^2,
\label{eq:comp}
\end{eqnarray}  
diverges as the critical point is approached.  The same applies to the right-hand side of this equation (\ref{eq:comp}) which involves the mean squared deviation,  $(\Delta N)^2 = \langle(N-\langle N\rangle)^2\rangle$, of the particle number from its average value $\langle N\rangle$.  Likewise,  fluctuations of the density (which plays the role of an LGT order parameter) become anomalously large at $T\rightarrow T_c$.  For very large systems in equilibrium such fluctuations are of little importance since $\Delta N/\langle N\rangle$ scales like $1/\sqrt{N}$.  However,  for systems with limited particle number $N$ statistical fluctuations can be significant even at temperatures considerably below $T_c$ where they can obscure signatures of the first-order phase transition,  an issue that is of relevance in the data analysis leading to the parameters (\ref{eq:critpoint}).  The probability for the occurrence of fluctuations is proportional to $\exp[-\Delta G/T]$ (Landau {\it et al.},1958; Goodman {\it et al.},1984),  where $\Delta G$ is the change of the Gibbs free energy,  $G = E - TS + PV$.  With $dG = -SdT+VdP$ this reduces to $VdP$ for an isothermal process,  and the change of $G$ between two points with densities $n_1$ and $n_2$ is:
\begin{eqnarray}
\Delta G = G(n_2) - G(n_1) = \int_1^2(V dP)_T = N\int_{n_1}^{n_2} {dn\over n}\left({\partial P\over \partial n}\right)_T~.
\end{eqnarray}
Close to the critical density $n_c$ the term $b n$ in (\ref{eq:vdw}) is about $1/3$ so that $P(T,n)$ can be expanded up to terms of order $n^2$.  It follows that
\begin{eqnarray}
\text{e}^{-\Delta G/T} \simeq \exp\left[-N\left(\ln{n_2\over n_1} +2B(T)(n_2-n_1)\right)\right]~,
\label{eq:fluct}
\end{eqnarray}
where $B(T) = b - a/T$ is the second virial coefficient of the VdW EoS, to be discussed in the following section. 
Consider now the gas-liquid phase coexistence region at $T < T_c$ near the critical point (see Figure \ref{Fig3}) where density fluctuations are expected to become relevant.  In this region the pressure as drawn by the VdW EoS involves  an interval of instability with $\partial P/\partial n < 0$.  This is not the pressure of the true equilibrium state which is given by the Maxwell-constructed horizontal dashed line in Fig. \ref{Fig3}.  Let us choose $n_1 = n_g$ and $n_2 = n_l$ at the start and end points of the coexistence region.  Anticipating from a forthcoming discussion (see section \ref{sec:critexp}) that a universal scaling rule relates $n_2-n_1 = n_l-n_g$ to $(1-T/T_c)^\beta$ near the critical point,  with the critical exponent $\beta$ (with a value $1/2$ in the mean field approximation,  in practice more like $1/3$),  we can now return to Eq.(\ref{eq:fluct}) and explore the role of density fluctuations more closely.  There is an evident competition between the limits $N\rightarrow\infty$ and $T\rightarrow T_c$.  For very large particle number $N$, approaching an infinite system,  fluctuations are suppressed as long as $T<T_c$.  However,  for a subsystem with finite $N$ the probability for the occurrence of fluctuations can be large.

Finally it is of interest to examine whether the empirically determined parameters $T_c, P_c$ and $n_c$ in Eq.\,(\ref{eq:critpoint}) of the nuclear LGT are reminiscent of a Van der Waals equation of state around the critical point.  One finds:
\begin{eqnarray}
b = {T_c\over 8\,P_c } = 7.2\pm 1.5~\text{fm}^3 = {1\over 3\,n_c} ~; ~~~\text{i.e.}~~n_c \simeq 0.05\pm 0.01~ \text{fm}^{-3}~.
\label{eq:bvalue}
\end{eqnarray}
Hence,  within their ranges of uncertainties,  the empirically deduced quantities that define the critical point of the nuclear LGT are indeed consistent with an equation of state of Van der Waals type.  Note that the critical density $n_c$  at $T_c$ is about $1/3$ of the nuclear matter density at equilibrium, $n_0\simeq 0.16$ fm$^{-3}$, at $T=0$. The observation that nuclear matter around $T_c$ appears to behave like a Van der Waals system is of some significance for gaining first qualitative insights into the nature of nuclear interactions.  In spin-isospin-symmetric nuclear matter at low density the attractive part of the interaction is known to be governed by scalar-isoscalar two-pion exchange in a way analogous to a VdW potential apart from additional screening by the non-zero pion mass.  The longest-range one-pion exchange,  on the other hand,  averages out at the Hertree level in leading order.  

\subsection{Virial expansion}

At low densities $n$ an expansion of the pressure $P$ in powers of $n$ is justified: 
\begin{eqnarray}
P(n,T) = n T\left[1+B(T)\,n + C(T)\,n^2 + ...\right]~.
\end{eqnarray}
This is the so-called virial expansion.  The leading correction to the EoS of an ideal gas is encoded in the second virial coefficient,  $B(T)$.  Expanding the EoS (\ref{eq:vdw}),  this coefficient is expressed in terms of the VdW parameters as:
\begin{eqnarray}
B(T) = b - {a\over T}~.
\end{eqnarray}
At $T=T_c$ it follows that $B(T_c) = -(19/8)\,b$.  With the ``empirical" value (\ref{eq:bvalue}) one finds 
\begin{eqnarray}
B(T_c) = -17.1\pm 3.6\,\text{fm}^3~,
\label{eq:Bcrit}
\end{eqnarray}
Consider now a system of particles interacting through a two-body potential ${\cal V}(r)$.  A standard textbook derivation (see e.g.  Landau {\it et al.} (1958),  Schwabl (2006)) starting from the grand canonical partition function relates the second virial coefficient to this potential as follows:
\begin{eqnarray}
B(T) = -2\pi\int_0^\infty dr\,r^2\left[\exp\left(-{{\cal V}(r)\over T}\right)-1\right] ~.
\label{eq:B(T)}
\end{eqnarray}
The value $B(T_c)$ displayed in (\ref{eq:Bcrit}),  even though it carries an uncertainty of about 20\%,  can now be used for orientation in order to constrain qualitative features of the underlying two-body potential ${\cal V}(r)$.

\subsection{Second virial coefficient and a schematic interaction}

The force behind self-bound spin-averaged nuclear matter with equal number of protons and neutrons involves a subtle balance between strong short-range repulsion and intermediate-range attraction.  At the low baryon densities $n < n_0$ characteristic of the liquid-gas coexistence region,  two-body interactions between nucleons dominate.  Three- und many-body interactions become important at higher densities.  While the short-range repulsive core of the interaction is commonly described in a phenomenologically parametrized form,  the longer-range attraction has a more explicit foundation within nuclear chiral dynamics,  the framework based on the spontaneously broken chiral symmetry of low-energy QCD,  and to be discussed more extensively in Chapter \ref{ch5}.  Nuclear chiral dynamics highlights the Goldstone boson character of the pion and consequently emphasizes its special role in the interaction between nucleons.  

\begin{figure}[t]
\centering
\includegraphics[width=.3\textwidth]{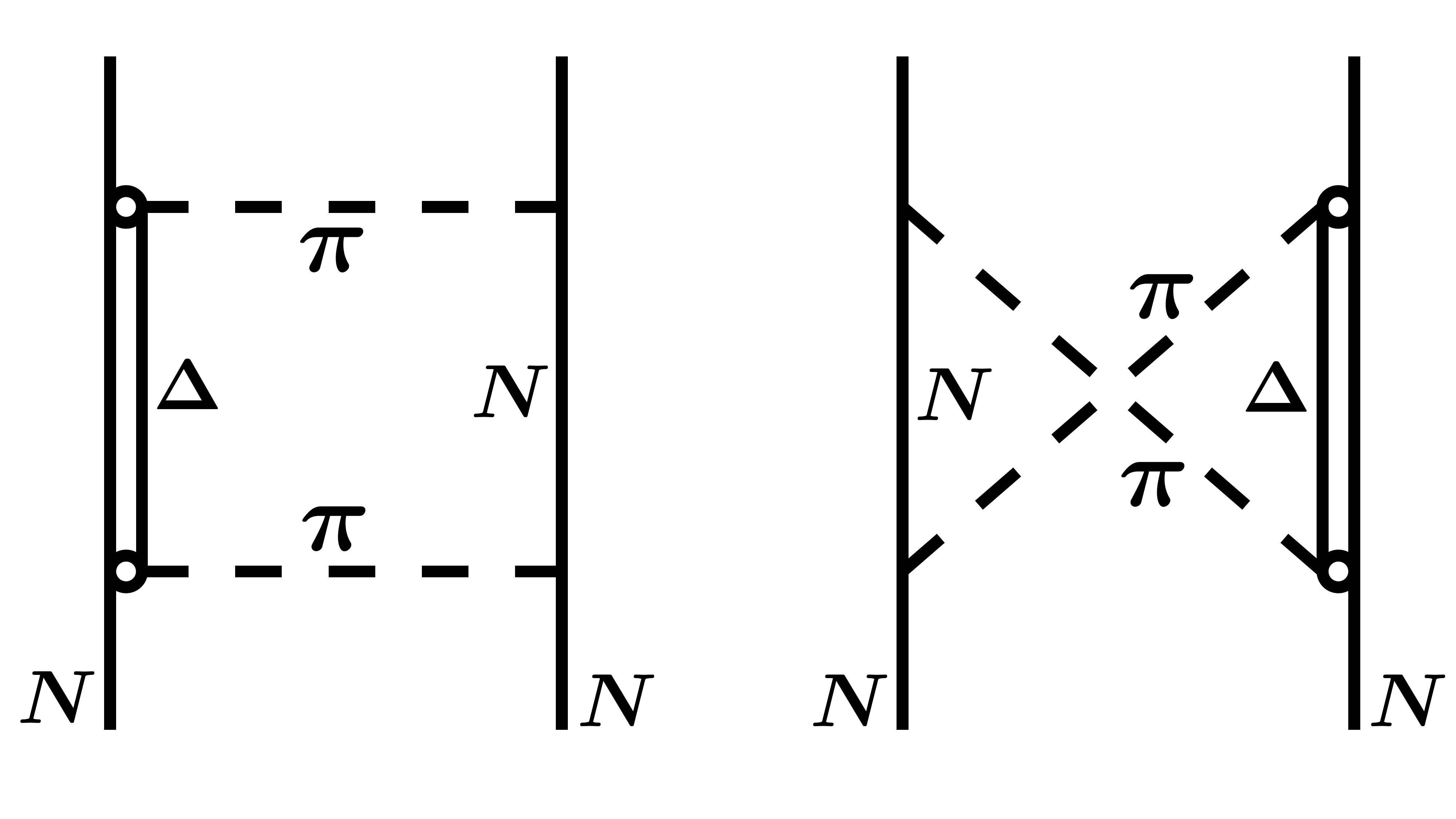}
\caption{Two-pion exchange interaction between nucleons with an intermediate $\Delta N$ state involving the virtual excitation of a $\Delta(1230)$ baryonic resonance. }
\label{Fig4}
\end{figure}

In spin-saturated nuclear matter the leading-order one-pion exchange interaction is not active at the Hartree (mean-field) level  because of spin averaging.  Two-pion exchange processes are at the origin of the attraction at intermediate and long distances.  Among those $2\pi$ exchange processes between two nucleons there is a dominant mechanism in which one of the exchanged pions turns a nucleon into an excited intermediate state,  the $\Delta(1230)$ resonance,  and then the second exchanged pion transfers the virtual excited system back to the two-nucleon ground state (see Fig.\,\ref{Fig4}).  This mechanism has qualitative features in common with a classical Van der Waals potential.  It is instructive here to combine it in a schematic ansatz together with a parametrized short-range repulsive interaction. 

A derivation of this VdW-like two-pion exchange potential with $\Delta N$ intermediate state results in the following form at intermediate and long distances (Kaiser {\it et al.},1998):
\begin{eqnarray}
{\cal V}_{2\pi}(r) = -W_S\,{\text{e}^{-2x}\over x^6}\,P(x)~,~~~~(x=m_\pi\,r)~,
\label{eq:2pipot}
\end{eqnarray}
with a polynomial in $x = m_\pi\,r$ :
\begin{eqnarray}
P(x)=1+2\,x+{5\over 3}\,x^2+{2\over 3}\,x^3+{1\over 6}\,x^4~.
\label{eq:poly}
\end{eqnarray}
The $1/r^6$ behavior in (\ref{eq:2pipot}),  typical of a VdW potential,  is recovered in the chiral limit of vanishing pion mass, $m_\pi\rightarrow 0$. 
The ``scalar" coupling strength $W_S$ is determined by a combination of known constants:
\begin{eqnarray}
W_S = {9\over 32\pi^2}\left({g_A\,m_\pi\over f_\pi}\right)^4{m_\pi^2\over\Delta}~,
\label{eq:poly}
\end{eqnarray}
with the nucleon axial vector coupling constant, $g_A = 1.276$,  the pion decay constant,  $f_\pi = 92.2$ MeV,  the pion mass,  $m_\pi = 139.6$ MeV and the $\Delta-N$ mass difference,  $\Delta = M_\Delta - M_N \simeq 291$ MeV. Using these empirical values we have $W_S \simeq 26.6$ MeV. The $x^{-6}$ singularity at short distance needs to be regularized.  As a convenient regularization one can choose:
\begin{eqnarray}
\tilde{P}(x) = \left[1- \exp\left(-{r^6\over d^6}\right)\right]\,P(x)~.
\label{eq:poly}
\end{eqnarray}
This introduces an additional distance scale $d$ at which repulsive effects begin to balance the attractive potential.
\begin{figure}[t]
\centering
\includegraphics[width=.48\textwidth]{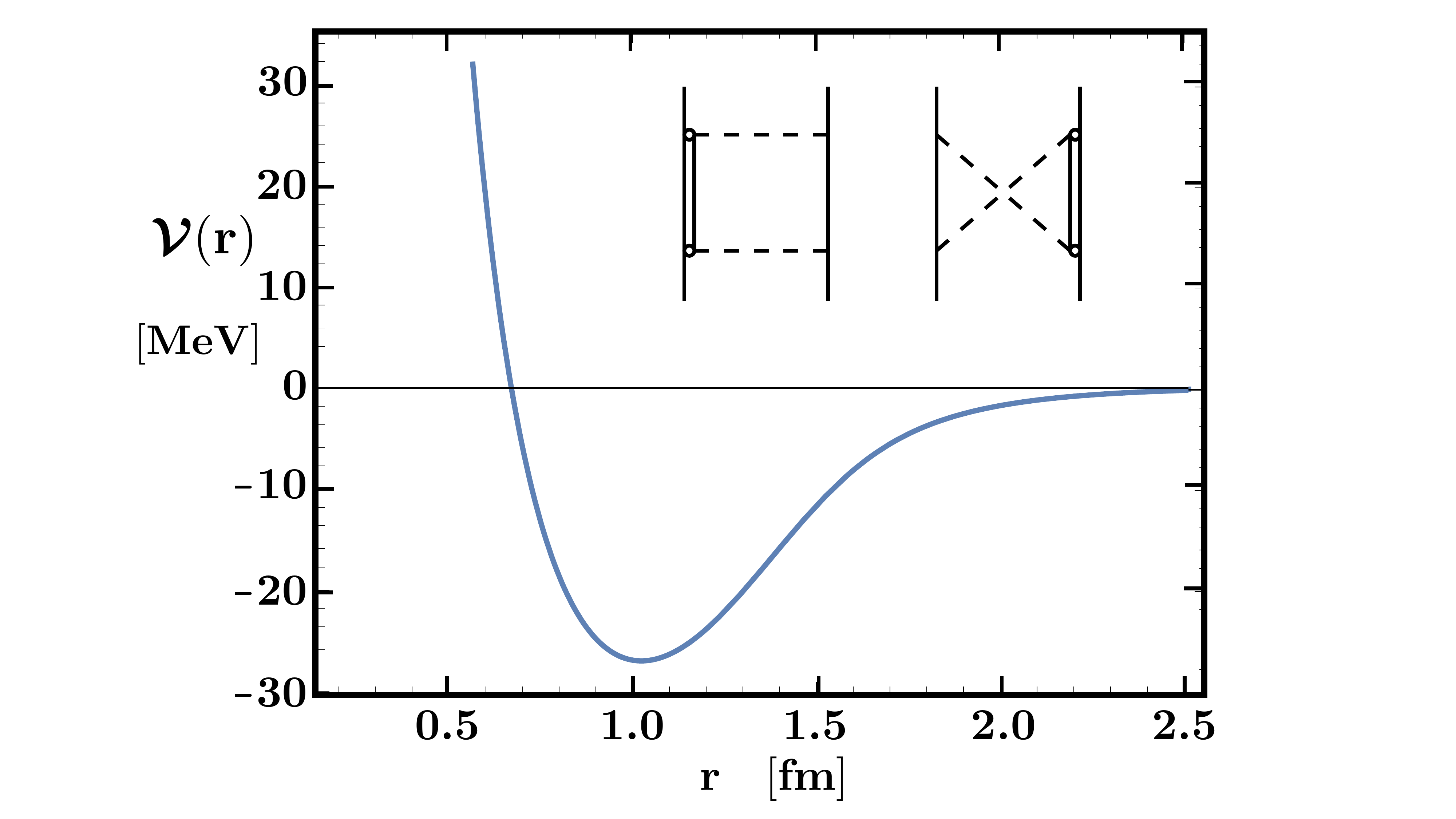}
\includegraphics[width=.48\textwidth]{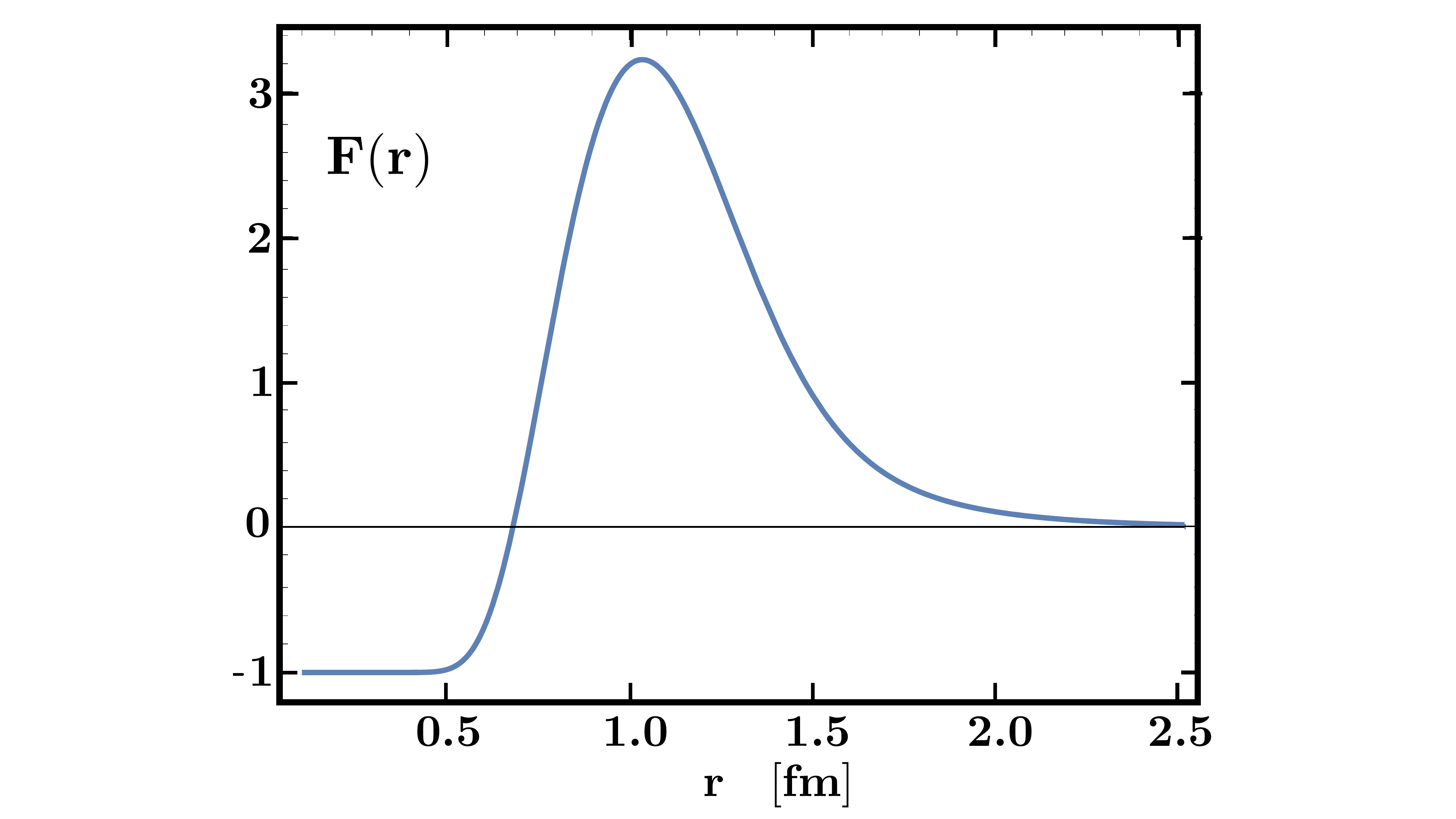}
\caption{Left panel: Schematic interaction potential (\ref{eq:potential}) that reproduces the empirical virial coefficient (\ref{eq:Bcrit}), $B(T_c)\simeq -17$ fm$^3$,  at the critical temperature $T_c = 17.9$ MeV.  The attractive part of the potential is based on two-pion exchange between nucleons with virtual excitation of the $\Delta(1230)$ baryonic resonance as sketched in Fig.\,\ref{Fig4} and indicated in the figure.  Right panel: Plot of the integrand $F(r) = \exp(-{\cal V}(r)/T_c) - 1$ of the virial coefficient $B(T_c)$ (see eq.\,(\ref{eq:integrand})).}
\label{Fig5}
\end{figure}

Combining the attractive part of the interaction with a repulsive Yukawa potential at short distance,  the following ansatz of a schematic potential can now serve as input into the second virial coefficient (\ref{eq:B(T)}),  to be compared with $B(T=T_c)$:
\begin{eqnarray}
{\cal V}(r) = W_V\,{\text{e}^{-m_V r}\over m_V r}-W_S\,{\text{e}^{-2x}\over x^6}\,\tilde{P}(x)~,
\label{eq:potential}
\end{eqnarray}
The short-range repulsive term is guided by boson exchange phenomenology of the nucleon-nucleon interaction.  This is commonly associated with the exchange of an isoscalar vector boson,  the omega meson being a natural candidate.  We use $m_V = 0.8$ GeV and $W_V \simeq 1.6$ GeV. 

The left panel of Fig.\,\ref{Fig5} presents the potential ${\cal V}(r)$ constrained to reproduce the median value (\ref{eq:Bcrit}) of the second virial coefficient at the critical temperature $T_c = 17.9$ MeV.  The regularization scale is tuned to $d= 1.31$ fm in order to arrive at this result.  It is also instructive to plot the integrand in $B(T_c) = -2\pi\int_0^\infty dr\, r^2 F(r)$,
\begin{eqnarray}
F(r) = \exp\left(-{{\cal V}(r)\over T_c}\right) - 1~, 
\label{eq:integrand}
\end{eqnarray}
shown in the right panel of Fig.\,\ref{Fig5}.  The plateau $F(r) = -1$ at small distances reflects the strong short-range repulsion.  An additional repulsive contribution from the regularization of the $2\pi$ exchange potential (the $\exp(-r^6/d^6)$ term in (\ref{eq:poly})) is hidden in this plot.  When taken together the combined repulsive effects integrate to a $b$-parameter of order $b \simeq (2\pi/3)d^3$.  (The empirical $b-$parameter (\ref{eq:bvalue}) suggests $d \simeq 1.5\pm 0.1$ fm.)

The schematic picture drawn here of the potential responsible for the nuclear liquid-gas phase transition is obviously much simplified.  It needs to be further extended towards detailed nuclear many-body calculations,  to be reviewed in the following chapters.  It is nonetheless worth noting that ${\cal V}(r)$ shown in Fig.\,\ref{Fig5} displays features quite similar to the $^1S_0$ and $^3S_1$ central $NN$ potentials reported from Lattice QCD computations (Aoki {\it et al.}, 2023).

\section{Thermostatics of nuclear matter}\label{ch4}

Many calculations of the EoS of nuclear matter at finite temperature have been performed starting from effective interactions that were adjusted to nuclear properties at zero temperature such as binding energies and matter radii.  A standard ansatz frequently used in this context is the Skyrme interaction.  Alternatively,  variational many-body calculations based on two-body potentials that reproduce nucleon-nucleon scattering phase shifts have also been applied.  Such approaches have also been extended to include three-body potentials fitted to properties of light nuclei.  We discuss representative examples of both these schemes and refer to a monograph (Hofmann, 2008) for a broader presentation.

\subsection{Thermal Hartree-Fock models}
An efficient method to compute the EoS of nuclear matter at finite temperature is the Hartree-Fock (HF) approach.  The free energy of the system is minimized in a self-consistent mean-field framework using a conveniently parametrized effective interaction (e.g.  Jaqaman {\it et al.},  1983).  A widely employed class of HF models connecting calculations for finite nuclei and infinite matter starts from a phenomenological effective interaction of the Skyrme type,  a density-dependent zero-range potential: 
\begin{eqnarray}
{\cal V}(\vec{r}\,) = t_0(1+x_0P_s)\,\delta^3(\vec{r}\,) + {t_1\over 2}(1+x_1P_s)\left[p^2\,\delta^3(\vec{r}\,)+\delta^3(\vec{r}\,)\,p^2\right]
+ t_2(1+x_2P_s)\left[\vec{p}\cdot\delta^3(\vec{r}\,)\,\vec{p}\right] + {t_3\over 6}(1+x_3P_s)\,n^\sigma\,\delta^3(\vec{r}\,)~,
\label{eq:effint}
\end{eqnarray}
where $\vec{r}= \vec{r}_1-\vec{r}_2$ is the relative distance between any two nucleons $(1,2)$,  and $\vec{p}=(\vec{\nabla}_1 - \vec{\nabla}_2)/2i$ is their relative momentum.  The operator $P_s={1\over 2}(1+\vec{\sigma}_1\cdot\vec{\sigma}_2)$ carries out spin exchange.  The spin-orbit term has been left out since its first-order contribution vanishes in a homogeneous system. The parameters $t_0,\dots,t_3$ and $x_0,\dots,x_3$ are adjusted to reproduce empirical nuclear properties.  The density-dependent term proportional to $n^\sigma$ with an adjustable exponent $\sigma$ is introduced to capture effects beyond two-body correlations. This Skyrme potential $\cal V$ is then used as input in the set of thermal HF equations (see box).  The nuclear matter EoS so obtained diplays a first-order liquid-gas phase transition, as expected.  

Various versions of Skyrme forces in HF computations have been employed and optimized over the years.  An early example (Sauer {\it et al.},  1976) of calculated pressure isotherms $P(T,n)$,  shown in Fig.\,\ref{Fig6}(a),  illustrates the so-called spinodal boundary that encloses the unstable region where ${\partial P\over \partial n}$, related to the compressibility,  is negative.  The region in which liquid and gas phases coexist,  obtained by a Maxwell construction,  includes this spinodal area.
The critical temperature $T_c = 17.9$ MeV and the location of the critical point in the pressure-density plane found in these calculations are quite notable as they compare very well with the more recent empirically deduced values (\ref{eq:critpoint}).

\begin{BoxTypeA}[box1]{Thermal Hartree-Fock approach}
\section*{Energy density}
Consider homogeneous (non-relativistic) nuclear matter with equal numbers of protons and neutrons (nucleon mass $M$) interacting through a two-body potential ${\cal V}$. The energy density of the system with density $n$ and temperature $T$ is
\begin{eqnarray}
\varepsilon(n,T) = \nu\int{d^3p\over (2\pi)^3}f(p,T){p^2\over 2M}+{\nu^2\over 2}\int{d^3p_1\over (2\pi)^3}\int{d^3p_2\over (2\pi)^3}f(p_1,T)f(p_2,T)\langle\vec{p}_1\vec{p}_2|\,{\cal V}\,|\vec{p}_1\vec{p}_2\rangle_a~,
\nonumber
\label{eq:epsilon}
\end{eqnarray}
with properly antisymmetrized matrix elements $\langle\vec{p}_1\vec{p}_2|\,{\cal V}\,|\vec{p}_1\vec{p}_2\rangle_a \equiv \langle\vec{p}_1\vec{p}_2|\,{\cal V}\,|\vec{p}_1\vec{p}_2-\vec{p}_2\vec{p}_1\rangle$ between plane-wave states.  Here $\nu$ is the spin-isospin degeneracy ($\nu = 4$ for isospin-symmetric nuclear matter). The density is given by
\begin{eqnarray}
n(T) = \nu\int{d^3p\over (2\pi)^3}f(p,T)~.
\nonumber
\label{eq:density}
\end{eqnarray}
In the thermal Hartree-Fock scheme the Fermi distribution function,
\begin{eqnarray}
f(p,T) = \left[1+\exp{{p^2\over 2M}+{\cal U}(p,T)-\mu\over T}\right]^{-1}~,
\nonumber
\label{eq:HFdist}
\end{eqnarray}
with baryon chemical potential $\mu$,  involves the self-consistently determined single-particle potential,
\begin{eqnarray}
{\cal U}(p,T) = \nu\int{d^3p'\over (2\pi)^3}f(p')\langle\vec{p}\vec{p}\,'|\,{\cal V}\,|\vec{p}\vec{p}\,'\rangle_a~.
\nonumber
\label{eq:HFpot}
\end{eqnarray}
\section*{Entropy density}
\begin{eqnarray}
s(n,T) = \nu\int{d^3p\over (2\pi)^3}\left[f(p,T)\ln f(p,T)+ \left(1-f(p,T)\right)\ln\left(1-f(p,T)\right)\right]~.
\nonumber
\label{eq:HFpot}
\end{eqnarray}
\section*{Free energy density and pressure}
\begin{eqnarray}
{\cal F} = \varepsilon - Ts~,~~~~~~P=\mu\,n -{\cal F}~,  ~~~\text{with chemical potential} ~~~~~\mu = {\partial {\cal F}\over\partial n}~.
\nonumber
\label{eq:HFpressure}
\end{eqnarray}
Given the approximate nature of the Hartree-Fock framework,  thermodynamic consistency requires confirming the validity of the relation $s = -{\partial{\cal F}\over\partial T}\big|_n$.
\end{BoxTypeA}

Multiple versions of Skyrme force parametrizations have appeared in the literature in the meantime. They differ in details such as the calculated nuclear compression modulus.  Given the broad spectrum of such effective interactions which are all more or less representative of nuclear ground state properties,  they permit estimates of the model dependence involved in the computations of the EoS and of calculated properties of the liquid-gas phase transition (Rios, 2010).  The critical temperatures obtained in those models cover a range $T_c \simeq 14 - 20$ MeV.

\subsection{Variational calculations}

An alternative scheme approaches the nuclear many-body problem by a variational calculation.  By construction,  this framework leads beyond the inherent mean-field level of Hartree-Fock computations with zero-range effective interactions.  It commonly uses as input a two-body potential that reproduces nucleon-nucleon scattering phase shifts and nuclear data, together with a phenomenological three-body interaction treated as a density-dependent part of the total effective potential.  The variational method applied to quantum fluids at finite temperature works with the concept of quasiparticle energies, $e(p,n,T)$,  which depend on momentum $p$ and density $n$ apart from the temperature $T$.  The internal energy density $\varepsilon(n,T)$ of the many-body system is given by the momentum space integral of the quasiparticle energies,  weighted by their occupation numbers in the Fermi sea.  The quasiparticle spectrum is often represented as $e(p,n,T) \simeq {p^2\over 2M^*(n,T)}$,  an approximation that was found to be adequate in actual applications.  The effective mass $M^*(n,T)$ is treated as a variational parameter at each $n$ and $T$.  The free energy density,  ${\cal F}(n,T) = \varepsilon(n,T) - T s(n,T)$, with the entropy density $s = - \partial {\cal F}/\partial T$,  is then minimized with respect to variations of $\varepsilon$ and $s$:
\begin{eqnarray}
\delta {\cal F}(n,T) = \delta \varepsilon(n,T) - T \delta s(n,T) ~.
\label{eq:var}
\end{eqnarray}
The variations $\delta \varepsilon$ and $\delta s$ include variations of the quasiparticle energies and of the quasiparticle occupation probabilties produced by changes of the chemical potential $\mu$ conserving the density $n$.

\begin{figure}[t]
\centering
\includegraphics[width=.7\textwidth]{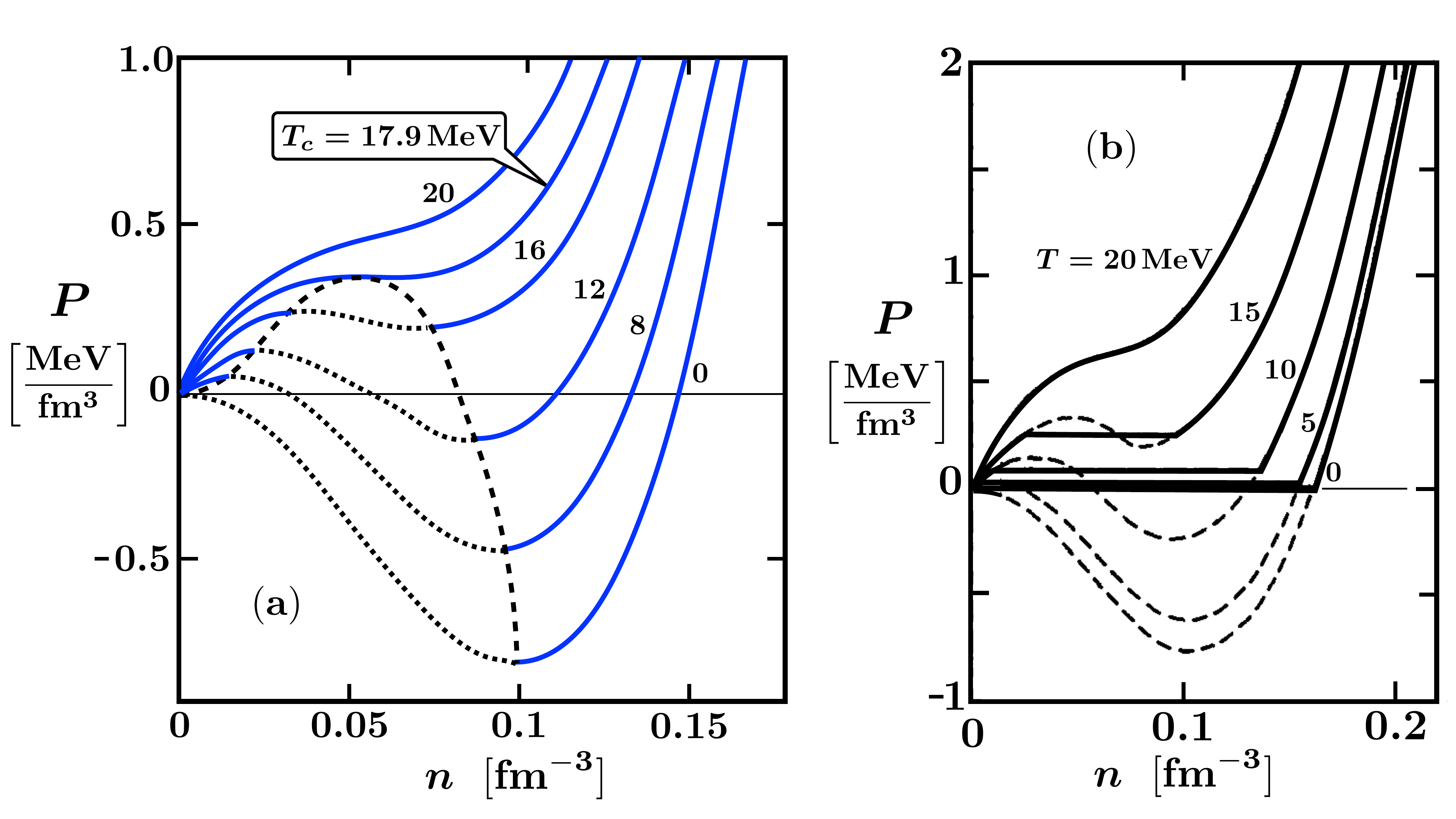}
\caption{Examples of EoS computations focusing on the first-order liquid-gas phase transition in symmetric nuclear matter: (a) Self-consistent Hartree calculation using a Skyrme effective interaction (adapted from Sauer et al.,1976); (b) Variational calculation using a two-body interaction that reproduces nucleon-nucleon scattering phase shifts,  plus a phenomenological three-body potential (the $\varv_{14} + TNI$ model); (figure adapted from Friedman et al.,1981). In both figures,  pressure isotherms are displayed as functions of density $n$,  with values of the temperature $T$ (in MeV) attached to each curve.}
\label{Fig6}
\end{figure}

An early successful example of such a variational approach to the nuclear LGT (Friedman {\it et al.},\,1981) is presented in Fig.\,\ref{Fig6}(b).  Horizontal lines of constant pressure in the isotherms at low temperature indicate regions of liquid-vapor two-phase equilibrium,  the phase coexistence region as it results from a Maxwell construction. The estimated critical temperature in this model is $T_c = 17.5\pm1.0$ MeV, consistent with the empirical $T_c$ as given in  (\ref{eq:critpoint}).

\subsection{Critical exponents}\label{sec:critexp}

As outlined in the previous sections,  mean-field models and variational approaches arrive at closely comparable results about the 1st order LGT and its critical point.  The calculated critical temperature,  pressure and density are at the same time consistent with their empirically deduced values.  These are in turn compatible with a prototypic Van der Waals EoS.  In a situation like this,  it is interesting to address certain universality properties of the phase transition near its critical point.  Such properties are described by so-called critical exponents,  parameters which encode universal characteristic behaviour in the vicinity of the critical point (Landau {\it et al.}, 1958; Kadanoff,  2000).  

In seeming contrast,  HF calculations using various different versions of Skyrme forces indicate a model-dependent spreading in the critical parameters of the phase transition.  It is therefore an issue whether the proposed universality close to the critical point is still realized.  An order parameter of the LGT is the density $n(P,T)$ which features a discontinuous change between the borders of the region where liquid and vapor phases coexist in equilibrium.  This discontinuity vanishes at the critical temperature $T_c$.  At this point the system undergoes a (continuous) second-order phase transition.  Consider the densities at the starting and end points of the phase coexistence region,  $n_l$ (for liquid) and $n_g$ (for gas),  respectively.  Close to the critical temperature universality arguments predict a behaviour:
\begin{eqnarray}
n_l - n_g \sim \left(1-{T\over T_c}\right)^\beta~,
\label{eq:discon}
\end{eqnarray}
Here $\beta$ is one of the critical exponents in question.  Another one,  $\gamma$,  is associated with the isothermal compressibilty close to the critical point:
\begin{eqnarray}
K_T= \left(n{\partial P\over\partial n}\right)_T^{-1} \sim \left(1-{T\over T_c}\right)^{-\gamma}~.
\label{eq:compr}
\end{eqnarray}
Yet another one, $\delta$,  relates the pressure as it approaches $P_c$ to the way the density approaches its critical value:
\begin{eqnarray}
|P-P_c| \sim |n-n_c|^\delta.
\label{eq:pressure}
\end{eqnarray}
The critical exponents are interconnected by scaling relations (Kadanoff {\it et al.}, 1967). For example:
\begin{eqnarray}
\gamma = \beta(\delta -1)~.
\end{eqnarray}
Landau's mean-field theory of phase transitions (Landau {et al.},1958) predicts the universal values $\beta = 1/2, \gamma = 1$ and $\delta = 3$ for systems in which thermal fluctuations can be ignored.

\begin{figure}[t]
\centering
\includegraphics[width=.7\textwidth]{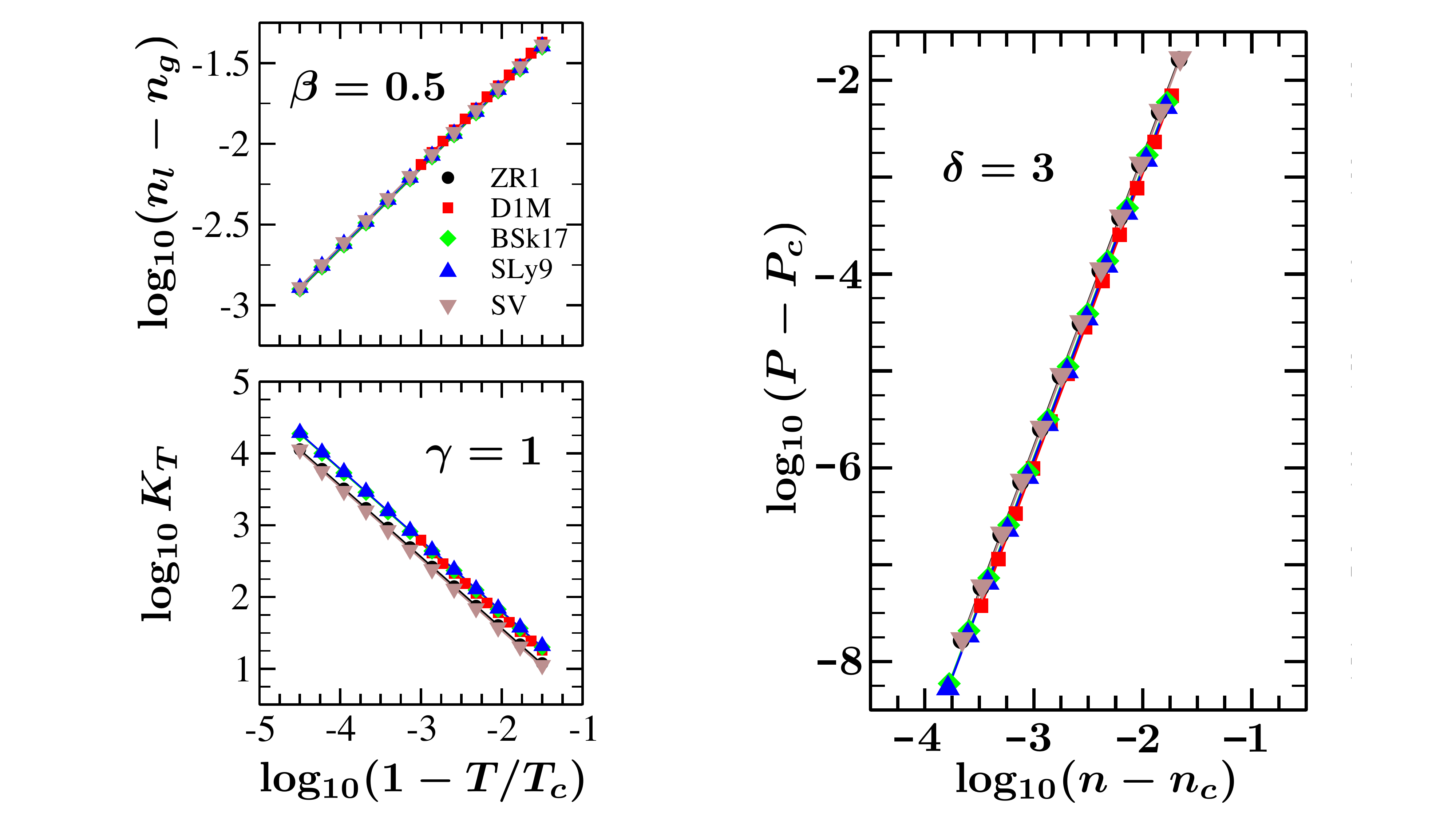}
\caption{Left panels: Logarithms of the density difference, $n_l - n_g$,  between liquid and gas phases and of the isothermal compressibility $K_T$ in the vicinity of the critical temperature $T_c$,  plotted against the logarithm of $1-T/T_c$.  Right panel: Plot of logarithms for pressure isotherms vs.  density in the vicinity of their critical values.  Data points from Hartree-Fock calculations using different effective interactions are lined up and fitted by linear regressions.  Slopes of the resulting lines determine the critical exponents $\beta, \gamma$ and $\delta$.  (Figures adapted from Rios (2010) - see this reference also for further details).}
\label{Fig7}
\end{figure}

A systematic study of universal critical behaviour has been carried out (Rios, 2010) using Hartree-Fock calculations with collections of different effective interactions.  The test is performed by taking logarithms of each of the relations (\ref{eq:discon}),  (\ref{eq:compr}) and (\ref{eq:pressure}).  Results are shown in Fig.\,\ref{Fig7}.  The double logarithmic plots display linear behaviours,  the slopes of which correspond to the values of the critical exponents.  These slopes turn out to be independent of the chosen interaction.  The self-consistent HF calculations give indeed critical exponents of the Landau mean-field type.  

The Hartree-Fock approximation to the nuclear many-body problem minimizes the thermodynamical potential by effectively reducing it to a one-body problem in a self-consistently determined mean field.  It may therefore not come as a suprise that the scale universality rules of Landau's mean-field theory of phase transitions are realized.  However this does not imply that mean-field methods are sufficient to describe the nuclear liquid-gas phase transition in all of its aspects.  Critical exponents apply only very close to the critical point.  At temperatures lower than $T_c$ the behaviour of the liquid-gas coexistence region is sensitive to details of the interaction.  Fluctuations beyond mean-field level can be important,  as it will become apparent also in subsequent analyses based on interactions that are derived from chiral effective field theory of low-energy QCD.

\section{Nuclear chiral thermodynamics}\label{ch5}
Following the pioneering work of Weinberg (1990, 1991),  Chiral Effective Field Theory (ChEFT) emerged and became established as a powerful tool to perform microscopic nuclear many-body calculations at densities below and around the nuclear equilibrium density, $n_0 \simeq 0.16$ fm$^{-3}$ (for a broad review see e.g.  Holt {\it et al.} (2016)).  This theory has its conceptual basis in the symmetry breaking pattern of Quantum Chromodynamics (QCD) at low energies.  Rather than with the quarks and gluons,  ChEFT is formulated in terms of nucleons and pions (and the $\Delta(1230)$ isobar excitation of the nucleon),  the relevant strong-interaction degrees of freedom that are effective at low energies and momenta.  Pions play a particularly pronounced role in this framework as (approximate) Nambu-Goldstone bosons (Nambu {\it et al.},  1961; Goldstone, 1961) of the spontaneously broken chiral symmetry of QCD in the low-energy limit.  For massless $up$ and $down$ quarks,  the pion emerges as a massless (soft) mode.  Its physical mass, $m_\pi \simeq 0.14$ GeV,  reflects the small but non-zero quark masses ($m_u \simeq 2$ MeV, $m_d\simeq 5$ MeV).  

\subsection{Liquid-gas transition in symmetric nuclear matter}

The ChEFT description of the nuclear many-body system makes use of two characteristic ``soft" scales: a) the pion mass $m_\pi$ is small compared to almost all other hadronic masses; b) nucleons at densities $n\lesssim n_0 \simeq 0.16$ fm$^{-3}$ behave non-relativistically,  i.e.  their momenta, $p\lesssim p_F = (3\pi^2 n/2)^{1/3}$,  are much smaller than their rest mass, $M\simeq 0.94$ GeV.  Indeed the Fermi momentum in symmetric nuclear matter at the equlibrium density $n_0$ is $p_F\simeq 0.26$ GeV $<< M$.  Given this separation of scales the approach to nuclear interactions and nuclear matter can be reliably organized using expansions in powers of $p, m_\pi <<\Lambda$ (Chiral Perturbation Theory,  ChPT),  where $\Lambda$ is a ``hard" scale\footnote{Calculations in ChPT often use the standard identification $\Lambda \equiv \Lambda_\chi = 4\pi f_\pi$, with the pion decay constant $f_\pi \simeq 92$ MeV.} of order 1 GeV.  This applies in particular to the density range $n < n_0$ relevant for the nuclear liquid-gas transition.

{\bf Chiral perturbation theory approach to thermal nuclear matter}. The systematic ChPT hierarchy of contributions to the nucleon-nucleon interaction (Bernard {\it et al.}, 1995; Epelbaum {\it et al.}, 2009) starts at leading order with one-pion exchange accompanied by momentum-independent contact terms representing unresolved short-distance dynamics.  At next-to-leading order ($NLO$) a first set of two-pion exchange processes enters together with contact terms quadratic in momenta.  Next-to-next-to-leading order ($N^2LO$) introduces further two-pion exchange mechanisms,  those which involve p-wave pion-nucleon scattering dominated by the $\Delta(1230)$. Three-body $NNN$ interactions also appear for the first time at $N^2LO$.  Order $N^3LO$ brings in exchanges of three pions in combination with other terms of increasing complexity,  and so forth.  The parameters (low-energy constants) associated with each order in this hierarchy are fixed to reproduce pion-nucleon and nucleon-nucleon scattering data.  Three-body terms are constrained by empirical information from few-nucleon systems ($^3H, \, ^3He$ and $^4He$).

These chiral two- and three-body interactions serve as input in perturbative many-body calculations (in-medium ChPT; Fritsch {\it et al.}, 2002; Fiorilla {\it et al.}, 2012).  At finite temperatures the appropriate thermodynamically consistent approach (Wellenhofer {\it et al}., 2014) starts from the free energy density, ${\cal F}(n,T)$, as a function of its natural variables, $n$ and $T$.  For isospin-symmetric nuclear matter with equal number of protons and neutrons the free energy density is written as a sum of convolution integrals in momentum space:
\begin{eqnarray}
{\cal F}(n,T) = 4\int_0^\infty dp\,{\cal K}_1(p)\,d(p,T)&+& \int_0^\infty dp_1\int_0^\infty dp_2\,{\cal K}_2(p_1,p_2)\,d(p_1,T)d(p_2,T)\nonumber \\ 
&+& \int_0^\infty dp_1\int_0^\infty dp_2\int_0^\infty dp_3\,{\cal K}_3(p_1,p_2,p_3)\,d(p_1,T)d(p_2,T)d(p_3,T)~.
\label{eq:fen}
\end{eqnarray}
Here 
\begin{eqnarray}
d(p,T) = {p^2\over 2\pi^2}\left[1+\exp{p^2/2M-\mu_0\over T}\right]^{-1}~
\end{eqnarray}
is the Fermi-Dirac distribution apart from a kinematical prefactor $p^2/2\pi^2$.  The effective one-body chemical potential $\mu_0(n,T)$ is determined by its relation to the density,
\begin{eqnarray}
n = 4\int_0^\infty dp\,d(p,T)~,
\end{eqnarray}
with the spin-isospin degeneracy factor 4 characteristic of symmetric nuclear matter.  At zero temperature,  $\mu_0(n,T=0)= p_F^2/2M$ and $n = 2p_F^2/3\pi^2$.  The ``true" chemical potential differs from $\mu_0$ and is given by $\mu = \partial{\cal F}/\partial n$.

The kernels ${\cal K}_i$ are grouped in a hierarchy of one-, two- and three-body terms.  The one-body kernel ${\cal K}_1$ represents the non-interacting nucleon-gas contribution to the free energy density:
\begin{eqnarray}
{\cal K}_1(p;n,T) = \mu_0(n,T)-{p^2\over 3M} -{p^4\over 8M^3}~.
\end{eqnarray}
The extra term $-p^4/8M^3$ ensures the correct energy per particle at $T=0$ up to order $p_F^4$ and gives an accurate approximation to the free energy density of a fully relativistic nucleon gas at the densities and temperatures of interest here.  The kernels ${\cal K}_2$ and ${\cal K}_3$ summarize the two- and three-body interactions including all one- and two-pion exchange terms up to the given chiral order. 

\begin{figure}[t]
\centering
\includegraphics[width=.5\textwidth]{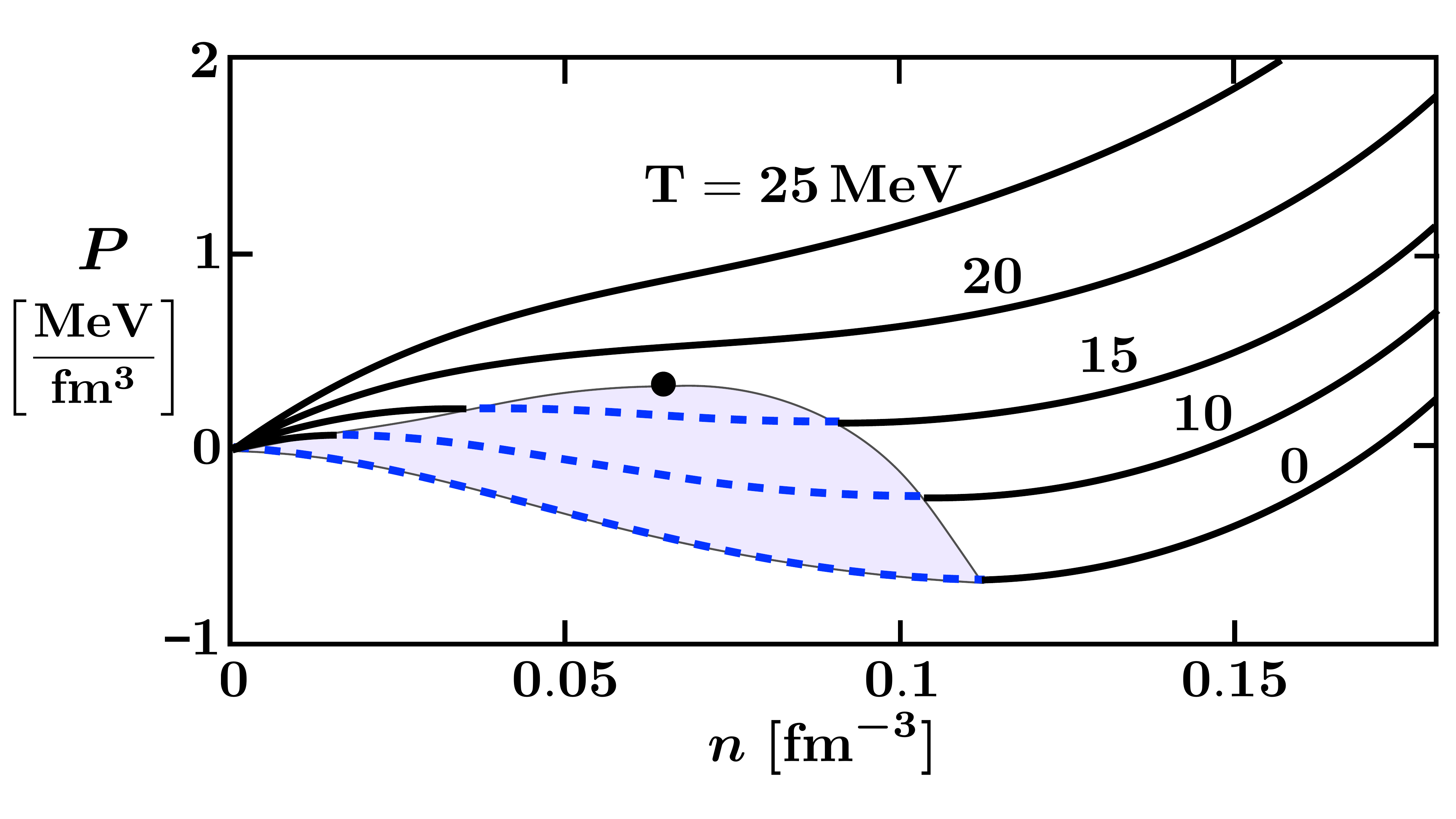}
\caption{Isotherms of the pressure $P(n,T)$ of isospin-symmetric nuclear matter calculated using many-body perturbation theory with $N^3LO$ chiral two-body interactions and $N^2LO$ three-body forces. The unstable spinodal region is marked explicitly.  The dot shows the location of the critical point.  (Adapted from Wellenhofer {\it et al.} (2015)).} 
\label{Fig8}
\end{figure}

The calculations of the free energy density ${\cal F}(n,T)$ are performed using the framework of Kohn-Luttinger-Ward many-body perturbation theory (Kohn {\it et al.}, 1960; Luttinger {\it et al.}, 1960). In practice these calculations are carried out to second order which is adequate at the low densities of interest.  The pressure $P(n,T)$ is derived as
\begin{eqnarray}
P(n,T) = n{\partial{\cal F}(n,T)\over \partial n} - {\cal F}(n,T)~.
\end{eqnarray}
Figure \ref{Fig8} shows a representative result for the EoS of symmetric nuclear matter with its first-oder liquid-gas phase transition and  a critical point located at temperature $T_c = 17.4$ MeV,  density $n_c=0.066$ fm$^{-3}$ and pressure $P_c = 0.33$ MeV/fm$^3$,  consistent with the empirical values (\ref{eq:critpoint}).  (A remaining weak cutoff dependence in the two-pion exchange interaction implies an uncertainty of about 3\% for these values (Wellenhofer {\it et al.}, 2015).)

{\bf Chiral thermodynamics and functional renormalization group}.  An alternative non-perturbative scheme features the combination of a chiral nucleon-meson model with the functional renormalization group (FRG) method (Berges {\it et al.}, 2002; Floerchinger {\it et al.}, 2012; Drews {\it et al.}, 2017).  The active baryonic degrees of freedom in this model are protons and neutrons collected in the nucleon field, $\Psi(x) = \left(\psi_p(x),\psi_n(x)\right)$ in terms of the Dirac fields of protons and neutrons.  The nucleons are sources of the chiral fields,  $\left(\sigma(x), \boldsymbol{\pi}(x)\right)$,  composed of the isovector-pseudoscalar pions $\boldsymbol{\pi}(x)$ as Nambu-Goldstone bosons and an isoscalar-scalar field $\sigma(x)$.  The starting point of this framework is the Lagrangian of the linear sigma model (Lee,  1972):
\begin{eqnarray}
{\cal L} = \bar{\Psi}\left[\text{i}\gamma^\mu\partial_\mu-g(\sigma-\text{i}\gamma_5\boldsymbol{\tau}\cdot\boldsymbol{\pi})\right]\Psi+{1\over 2}\left[(\partial^\mu\sigma)(\partial_\mu\sigma) + (\partial^\mu\boldsymbol{\pi})\cdot(\partial_\mu\boldsymbol{\pi})\right] - {\cal U}(\sigma,\boldsymbol{\pi})~.
\label{eq:sigmamodel}
\end{eqnarray}
It includes Yukawa couplings of the nucleons to the chiral fields with a common coupling constant $g$.  The expectation value of the sigma field, $\langle\sigma\rangle\equiv\sigma_0$,  acts as an order parameter of spontaneously broken chiral symmetry.  It is normalized in the vacuum as $\sigma_0 = f_\pi$,  with the empirical pion decay constant $f_\pi = 92.2$ MeV.  Its evolution in a thermal nuclear many-body system can be associated with the in-medium pion decay constant,  $f_\pi^*(n,T)$.  The nucleon mass in this model is dynamically linked to the sigma mean field as $M = g\langle\sigma\rangle$,  its vacuum value being $M_0 = gf_\pi$,  while its in-medium evolution follows the behaviour of $f_\pi^*(n,T)$. The bosonic potential ${\cal U}(\sigma,\boldsymbol{\pi})$ has a chiral invariant part written in terms of powers of $\sigma^2 + \boldsymbol{\pi}^2$, plus a term proportional to $\sigma$ that breaks chiral symmetry explicitly and reflects the small non-zero average quark mass $(m_u+m_d)/2$ in QCD.  The minimum of the so constructed $\cal U$  is located at $\sigma = \sigma_0 = f_\pi$ for $\boldsymbol{\pi}=0$.  

Applications to nuclear many-body systems require adding a piece to (\ref{eq:sigmamodel}) that encodes the information about the strong short-range repulsion in the nucleon-nucleon interaction.  This is conveniently done, maintaining chiral symmetry,  by introducing phenomenological isoscalar and isovector vector fields coupled to the nucleons with parameters tuned to reproduce nuclear matter ground state properties.
Finite temperature together with chemical potentials for protons and neutrons are implemented using the Matsubara formalism: Minkowskian space-time $x^\mu =(t,\boldsymbol{x})$ is rotated to Euclidean space,  replacing $t\rightarrow-\text{i}\tau$.  The $\tau$-dimension is compactified to a circle such that $\tau$ is restricted to $[0,\beta]$ with the inverse temperature, $\beta=1/T$.  Boson and fermion fields are periodic or antiperiodic,  respectively,  under $\tau\rightarrow\tau+\beta$.  These procedures transform the Minkowski-space action, ${\cal S} = \int d^4x\,{\cal L}$,  into the Euclidean action,  ${\cal S}_E = \int_0^\beta d\tau\int d^3x\,{\cal L}_E$,  which is the basis for computing the grand canonical partition function.

The FRG method applied to nuclear thermodynamics now proceeds as follows\footnote{for a more detailed presentation see e.g.  Drews {\it et al.} (2017)}.  The Euclidean action ${\cal S}_E$ of the chiral nucleon-meson model works in the domain of spontaneously broken chiral symmetry,  i.e.  within a momentum range below the characteristic symmetry breaking scale,  $\Lambda_\chi = 4\pi\,f_\pi\sim 1\,\text{GeV}$.  This scale is taken as the ``ultraviolet" initialization of a renormalization-group flow equation (Wetterich, 1993) that describes the evolution of the action down to low-energy ``infrared" scales characterizing the nuclear many-body problem at Fermi momenta $p_F<<\Lambda_\chi$.  Solving the FRG flow equations implies iterating loops of the propagating ``active" degrees of freedom (pions in particular) to all orders.  In essence this non-perturbative scheme permits a systematic treatment of fluctuations and thus goes beyond the mean-field approximation.  In practice these quantum fluctuations are represented by higher order terms of the effective potential ${\cal U}(\sigma,\boldsymbol{\pi})$ with coefficients depending on temperature $T$ and chemical potentials $\mu_p, \mu_n$.  At the same time this procedure should match the perturbative ChEFT approach at the low-energy domain of its applicability.  It is thus of interest to compare results for the nuclear liquid-gas phase transition obtained with both the in-medium ChPT and FRG schemes.

\begin{figure}[t]
\centering
\includegraphics[width=.5\textwidth]{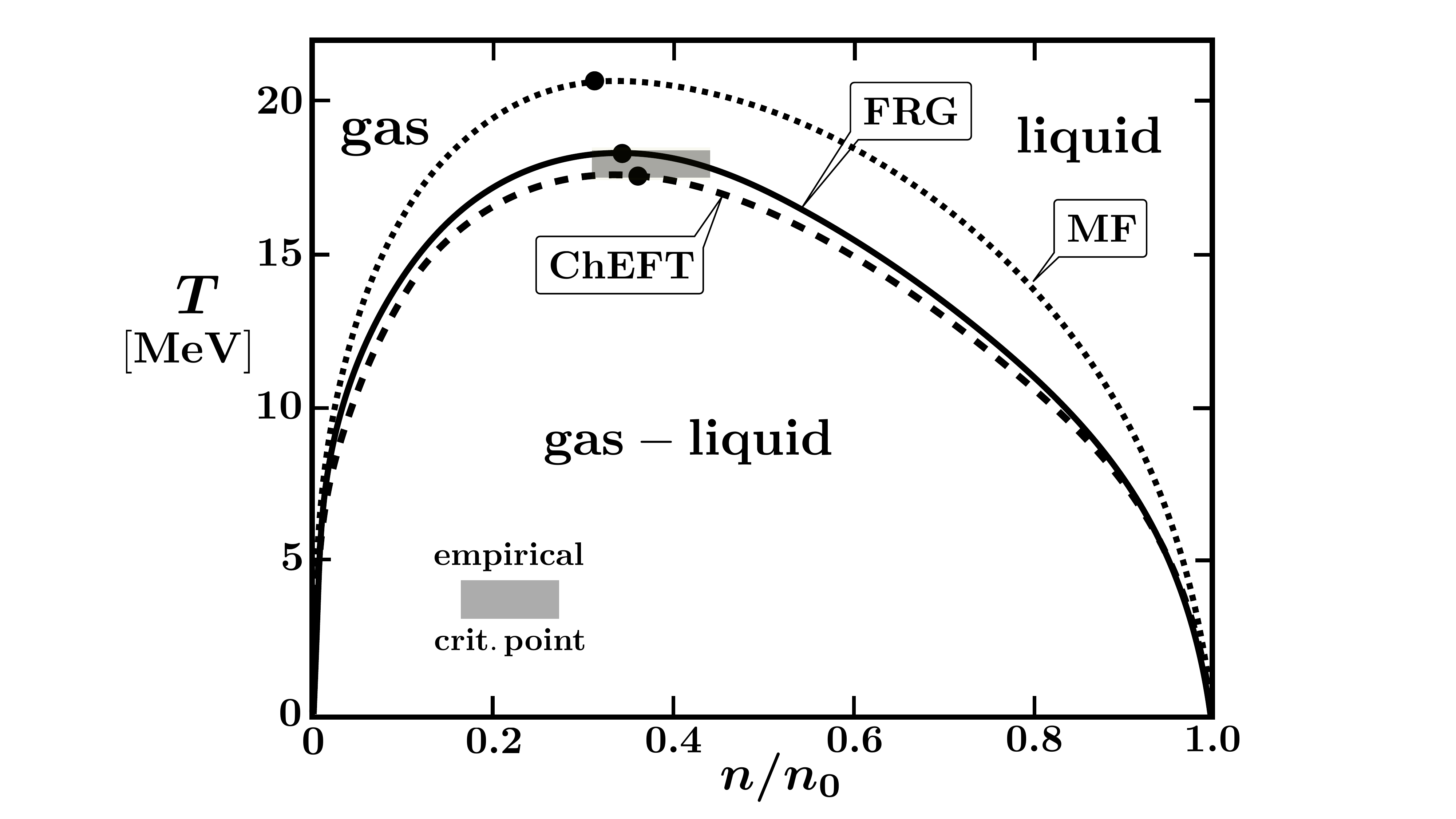}
\caption{Coexistence region of liquid and gas phases in isospin-symmetric nuclear matter.  Solid line (FRG): results of functional renormalization group approach using a chiral nucleon-meson model (Drews  {\it et al.}, 2017); dashed line (ChEFT): in-medium ChPT calculation using $N^3LO$ chiral $NN$ potentials together with $N^2LO$ three-body  interactions in many-body perturbation theory (Wellenhofer {\it et al.}, 2015); dotted line (MF): mean-field approximation to the FRG computation.  Dots indicate the locations of the critical point found in these calculations. The empirical range (Elliot {\it et al.}, 2013) of the critical point  $(n_c, T_c)$ is indicated by the shaded rectangle. The baryon density along the horizontal axis is given in units of nuclear matter equilibrium density,  $n_0 = 0.16$ fm$^{-3}$. }
\label{Fig9}
\end{figure}

The FRG calculations with the previously described chiral nucleon-meson model as input have been carried out both in mean-field (MF) approximation,  with the chiral fields $(\sigma, \boldsymbol{\pi})$ replaced by $\langle\sigma\rangle$,  and in the full setup including fluctuations beyond MF.  It is instructive  to compare these results with those of the perturbative ChEFT calculation mentioned before.  Representative examples are shown in Figure\,\ref{Fig9} where calculated boundary curves of the liquid-gas coexistence region of the LGT in isospin-symmetric nuclear matter are displayed.  The non-perturbative FRG and perturbative ChEFT results are indeed found to be close together.  The estimated critical temperature in the FRG calculation is $T_c = 18.3$ MeV,  still within the empirical range.  The comparison with a mean-field approximated FRG result emphasizes that, in particular,  pionic fluctuations are important.  Without these ingredients the critical temperature would be significantly overestimated in such chiral models. 

\subsection{Asymmetric nuclear matter: isospin dependence of the liquid-gas phase transition}

Isospin-symmetric nuclear matter is a self-bound system,  while pure neutron matter is unbound.  Isospin-asymmetric nuclear matter is characterized by unequal densities of protons ($n_p$) and neutrons ($n_n$),  with total baryon density $n=n_p+n_n$.   It is instructive to investigate how the pattern of the liquid-gas phase transition changes (Ducoin {\it et al.}, 2006) as the fraction of protons,
\begin{eqnarray}
x_p= {n_p\over n_p+n_n}~,
\label{eq:pfraction}
\end{eqnarray}
is reduced systematically from $x_p = 0.5$ for symmetric nuclear matter to neutron-rich matter as it is realized, for example,  in neutron stars.  This is where the isospin dependence of the nuclear interactions enters. The pion with its isospin $I=1$ charge triplet $\pi^+, \pi^0, \pi^-$ coupled to the isospin $I=1/2$ doublet $\Psi = \left(\psi_p,\psi_n\right)$ of proton and neutron plays a prominent role.  In ChEFT at leading orders one- and two-pion exchange processes generate the isospin-dependent parts of the nucleon-nucleon interaction at long and intermediate ranges.  Short-range mechanisms,  often interpreted as related to the exchange of $\rho$ mesons,  add to this picture. 

\begin{figure}[t]
\centering
\includegraphics[width=.5\textwidth]{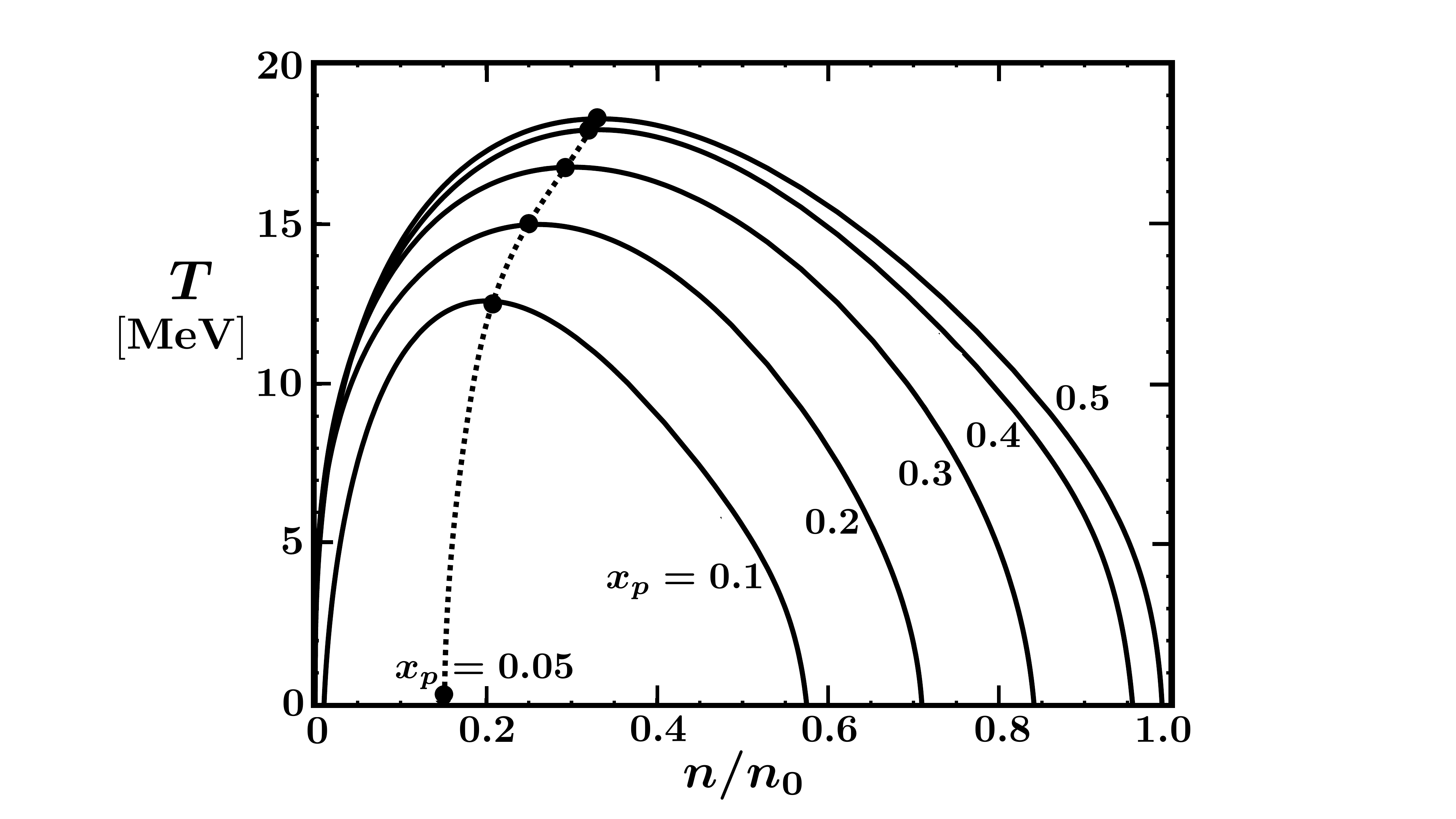}
\caption{Envelopes of liquid-gas coexistence regions in isospin-asymmetric nuclear matter for different proton fractions $x_p$.  Dots mark critical points of the first-order liquid-gas phase transition for each $x_p$.  The trajectory of critical points as a function of $x_p$ is indicated by the dotted line.  The baryon density along the horizontal axis is given in units of the equilibrium density,  $n_0 = 0.16$ fm$^{-3}$, of symmetric nuclear matter. Results are from calculations in a chiral nucleon-meson model combined with FRG methods (Drews {\it et al.}, 2017).}
\label{Fig10}
\end{figure}

The free energy density of homogeneous isospin-asymmetric nuclear matter can be expressed in terms of an expansion in the asymmetry parameter $\delta = (n_n-n_p)/(n_n+n_p) = 1-2x_p$:
\begin{eqnarray}
{\cal F}(n,T;\delta) = {\cal F}(n,T;\delta=0) + {\cal F}_{\text{sym}}(n,T)\,\delta^2 + {\cal O}(\delta^4\log|\delta|)~.
\label{eq:asym1}
\end{eqnarray}
The first ($\delta = 0$) term corresponds to symmetric nuclear matter,  while the ${\cal F}_{\text{sym}}$ part includes the changes of the free energy density related to the neutron-proton asymmetry.  It has been demonstrated in various microscopic many-body calculations that the isospin-asymmetry dependence of ${\cal F}(n,T;\delta)$ at $T=0$ is quadratic in $\delta$ to good accuracy.  At finite temperature the non-interacting Fermi gas contribution to the free energy density introduces higher than quadratic powers of $\delta$.  However,  in the temperature and density range relevant for the LGT it turns out that the quadratic parametrization in $\delta$ is still a valid approximation (Wellenhofer {\it et al.}, 2015). The asymmetry term ${\cal F}_{\text{sym}}$ is then given by the difference of free energy densities for pure neutron matter and symmetric nuclear matter:
\begin{eqnarray}
{\cal F}_{\text{sym}}(n,T) = {\cal F}(n,T;\delta = 1) - {\cal F}(n,T;\delta=0) ~.
\label{eq:asym2}
\end{eqnarray}
The pressure at each given value of the asymmetry $\delta$ is 
\begin{eqnarray}
P(n,T;\delta) = n{\partial{\cal F}(n,T;\delta)\over\partial n} - {\cal F}(n,T;\delta)~.
\label{eq:asymP}
\end{eqnarray}
The characteristic features of the nuclear LGT display a strong dependence on the proton-neutron composition in asymmetric matter.  As an example,  the liquid-gas coexistence region with its critical end point shrinks systematically as the proton fraction decreases from $x_p = 0.5$ towards neutron-rich matter.  At $x_p \simeq 0.05$ the LGT disappears as the system becomes unbound and proceeds towards pure neutron matter.
Figure \ref{Fig10} shows this pattern for the example of FRG calculations using a chiral nucleon-meson model.

\section{Nuclear liquid-gas transition and phase diagram of strongly interacting matter}
\label{ch6}

The nuclear liquid-gas phase transition is embedded in the phase diagram of QCD.  While the density and temperature range of the LGT is well separated from the confinement-deconfinement crossover transition at temperatures $T\sim 150$ MeV,  it is nonetheless of some interest to locate the LGT within the framework of spontaneous chiral symmetry breaking,  one of the prominent low-energy properties of QCD.

\subsection{Liquid-gas transition in the context of chiral symmetry breaking}

A possible order parameter encoding the pattern of spontaneously broken chiral symmetry in strongly interacting matter is the pion decay constant.  It is related to the lifetime of the charged pion with respect to its weak decay into muon and muonic antineutrino.  Its vacuum value,  $f_\pi = 92.2$ MeV,  changes in a dense and hot medium and splits into a temporal part,  $f_\pi^*(n,T)$,  and a spatial part which is of no interest here.  By its decreasing tendency at non-zero temperatures and baryon densities,  the in-medium pion decay constant, $f_\pi^*(n,T)$,  acts as a measure indicating the approach to the restoration of chiral symmetry at high temperatures and/or densities.

In models based on chiral symmetry,  such as the linear sigma model which is at the origin of Eq.\,(\ref{eq:sigmamodel}),  the in-medium pion decay constant is identified with the expectation value of the sigma field,
\begin{eqnarray}
\langle\sigma\rangle(n,T) = f_\pi^*(n,T)~.
\label{eq:orderpara}
\end{eqnarray}
Alternatively,  the squared pion decay constant appears in an equation (Gell-Mann {\it et al.}, 1968) establishing its connection with the chiral (quark) condensate,  $\langle\bar{q}q\rangle$,  that also acts as an indicator of spontaneous chiral symmetry breaking in QCD.

\begin{figure}[t]
\centering
\includegraphics[width=.5\textwidth]{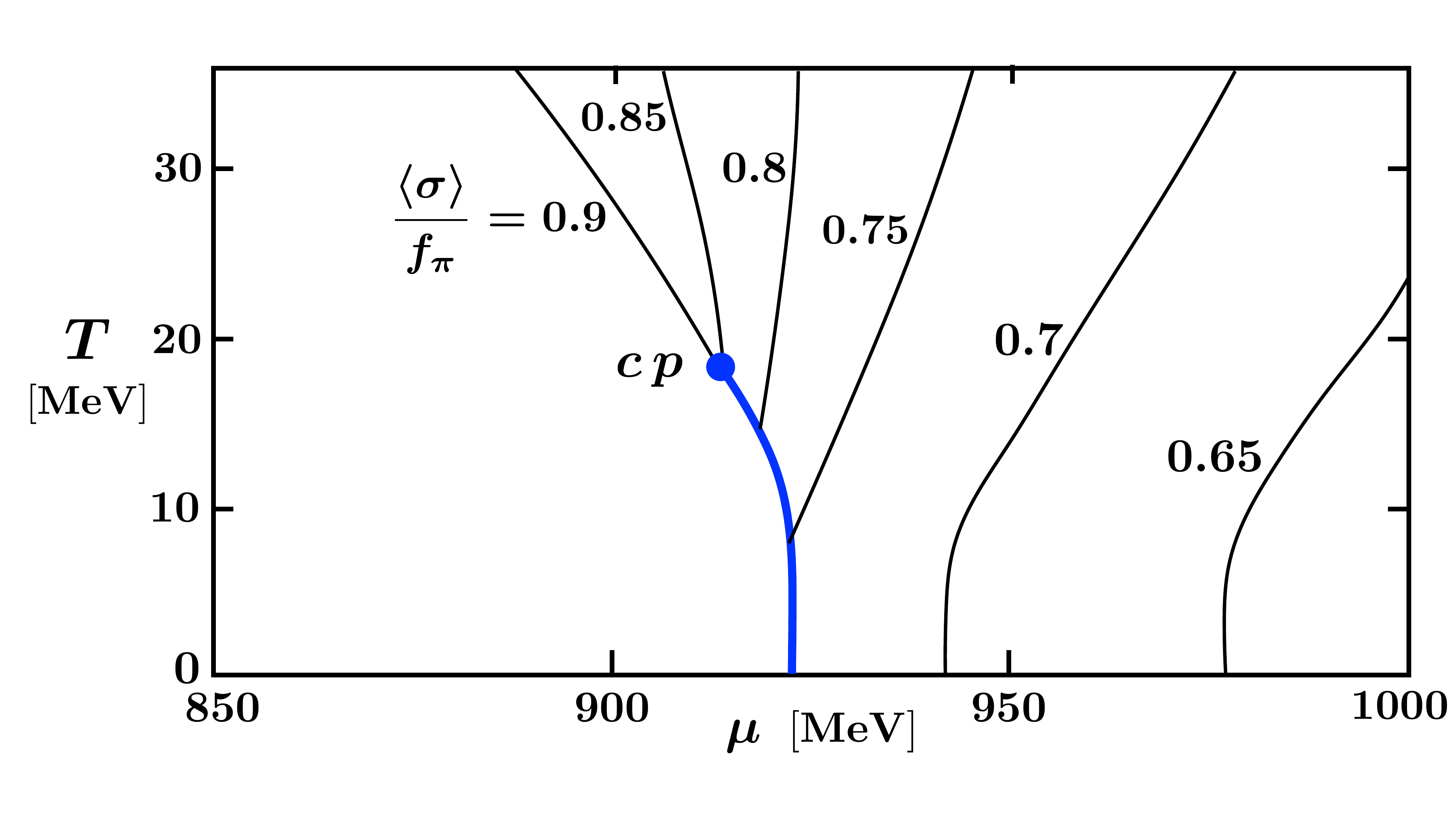}
\caption{Temperature vs.  baryon chemical potential in the neighborhood of the liquid-gas first order phase transition in isospin-symmetric nuclear matter.  The blue phase discontinuity line terminates at the critical end point (cp).  Surrounding thin curves display contours of the decreasing chiral order parameter,  $\langle\sigma\rangle/f_\pi$.  Calculations are performed in the chiral nucleon-meson model combined with FRG methods.  (Adapted from Drews {\it et al.},   2017).}
\label{Fig11}
\end{figure}

For the understanding of the nuclear liquid-gas phase transition as a low-energy phenomenon it is important to confirm that the LGT is entirely located well within the domain of spontaneously broken chiral symmetry where the relevant effective degrees of freedom are indeed nucleons with a mass close to their values in vacuum.  This is conveniently demonstrated in a phase diagram with temperature $T$ and baryon chemical potential $\mu$ as coordinates.  We recall that the chemical potential is given in terms of the free energy density ${\cal F}(n,T)$ by
\begin{eqnarray}
\mu(n,T) = {\partial{\cal F}(n,T)\over\partial n}~.
\label{eq:chempot}
\end{eqnarray}
In the $\mu-T$ phase diagram the first-order liquid gas transition in symmetric nuclear matter shows up as a discontinuity line starting at $T=0$ and the chemical potential $\mu_0 = M-B \simeq 923$ MeV associated with the nuclear matter saturation point (the minimum of the energy per particle),  where $M$ is the free nucleon mass and $B \simeq 16$ MeV the binding energy per nucleon.  The critical line terminates at its end point in a second-order phase transition.  It is then informative to follow contours of constant chiral order parameter, $f_\pi^*(\mu,T)$,  in the vicinity of the LGT line as shown in Figure\,\ref{Fig11}.  Evidently the liquid-gas transition is indeed far separated from regions of the phase diagram where one expects the restoration of chiral symmetry to occur.  Similar conclusions are also drawn in alternative chiral models, such as the parity doublet model in which the nucleon mass is only partially correlated with the chiral order parameter (Eser {\it et al.}, 2024).

\subsection{Liquid-gas transition and chemical freeze-out}

High-energy collisions of nuclei trigger a sequence of events from compression, forming hot and possibly dense matter for a short time, through expansion and cooling to a stage at which finally non-interacting hadrons and nuclei emerge.  This final stage is referred to as chemical freeze-out.  It is explored by systematically tracing produced particles in such collisions and analysing the characteristic temperatures and chemical potentials at which these particles finally appear as decoupled from one another.  In a $\mu-T$ phase diagram these chemical freeze-out points line up at temperatures far above and baryon chemical potentials well below the liquid-gas transition (Braun-Munzinger {\it et al.},  2026, and refs. therein).

\begin{figure}[t]
\centering
\includegraphics[width=.6\textwidth]{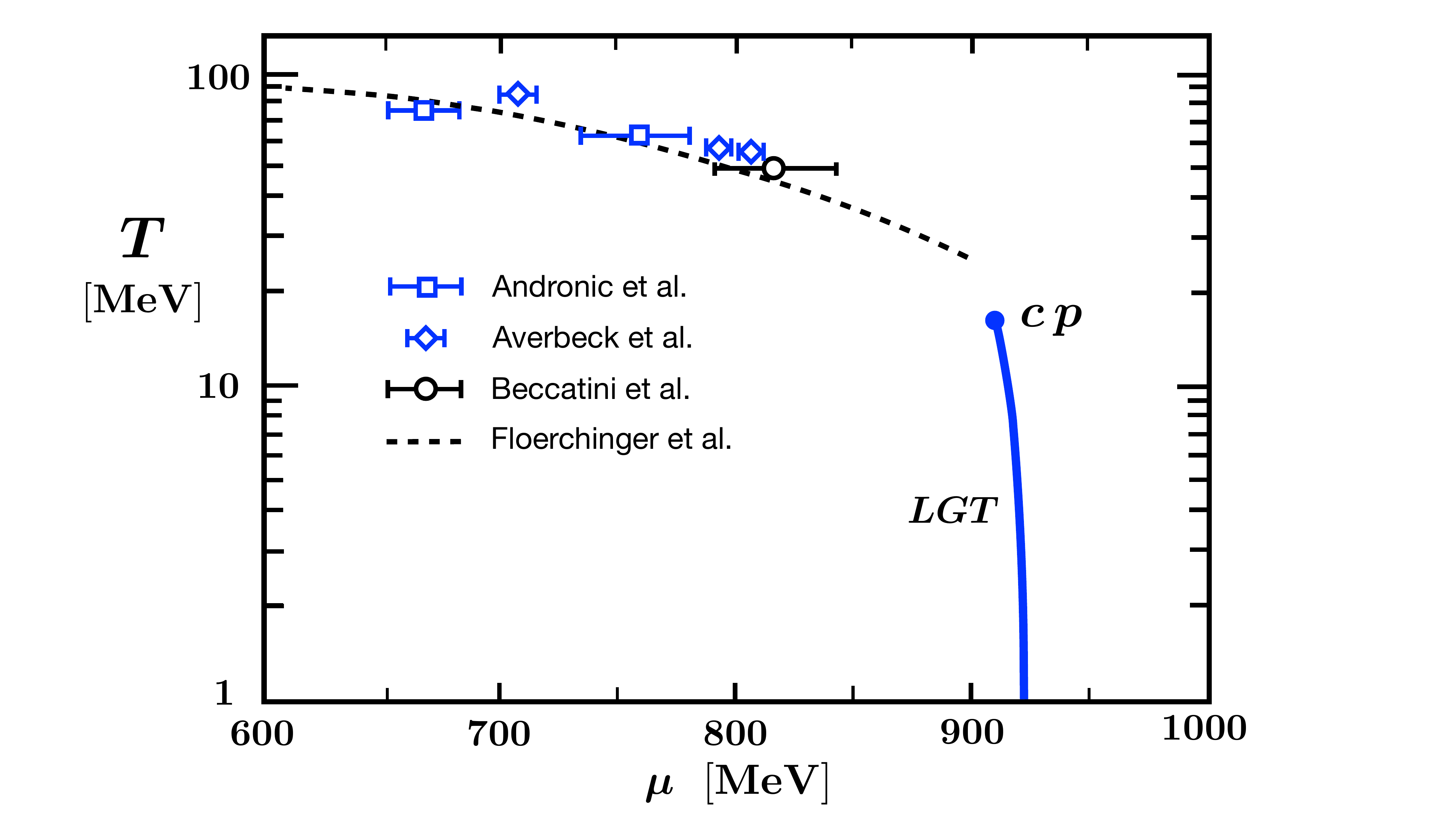}
\caption{Phase diagram (temperature vs. baryon chemical potential) showing the critical line (LGT) of the first-order liquid-gas phase transition (FRG calculation; Drews {\it et al.}, 2017) terminating in the critical point (cp),  in relation to chemical freeze-out points deduced from relativistic heavy-ion collision data (Andronic {\it et al.},  2018; Averbeck {\it et al.}, 2003; Beccatini {\it et al.}, 2001).  Dashed curve: line of constant (low) baryon density with $n=0.15\,n_0$ (Floerchinger {\it et al.}, 2012).}
\label{Fig12}
\end{figure}

The systematics of empirical chemical freeze-out data in the phase diagram resembles the deconfinement and chiral crossover transition bands deduced and extrapolated from Lattice QCD.  These bands extend over a range of low baryon chemical potentials and temperatures $160\,\text{MeV}\gtrsim T\gtrsim 140\, \text{MeV}$,  starting from the crossover at $T\simeq 157$ MeV and zero chemical potential as it is widely agreed upon (Braun-Munzinger {\it et al.},  2026).  Several chemical freeze-out points are also recorded at lower temperatures,  $T < 100$ MeV,  and larger baryon chemical potential reaching out to $\mu \simeq 650 - 850$ MeV,  as shown in Fig.\,\ref{Fig12}.  This data set,  which one might envisage as continuing toward the LGT region further down in temperature,  has sometimes been tentatively discussed as a possible extension of the QCD crossover transition to lower $T$ and larger $\mu$.  However,  this pattern and its position relative to the LGT critical line do not suggest such an interpretation.  In fact the freeze-out data displayed in Fig.\,\ref{Fig12} turn out all to be located close to a curve of constant, very low baryon density (a small fraction of the nuclear equilibrium density $n_0$,  see Floerchinger {\it et al.}, 2012).  It has thus been pointed out that no phase transition other than the LGT is expected in the range $T< 100$ MeV and $0.65\,\text{GeV}\lesssim\mu\lesssim 0.95\,\text{GeV}$ covered by Figure \ref{Fig12}.

\section{Summary}\label{summary}

\begin{itemize}
\item{The systematic analysis of multifragmentation data obtained from collisions between nuclei suggests that isospin-symmetric nuclear matter,  as an idealized infinite system with equal numbers of protons and neutrons (and Coulomb interaction switched off),  undergoes a first-order phase transition from a liquid to a vapor state.}

\item{This phase transition involves a liquid-gas coexistence region which terminates in a second-order phase transition at a critical end point. The position of this critical point is emprically constrained at a temperature $T_c \simeq 18$ MeV with an uncertainty of less than $3\%$.  The values of pressure and baryon density at this point are $P_c \simeq 0.3$ MeV/fm$^3$ and $n_c \simeq n_0/3$ (with $n_0$ the equilibrium density of nuclear matter).  These latter quantities are determined less accurately than the critical temperature,  with an uncertainty of about $20\%$.}

\item{The nuclear liquid-gas transition displays properties that can be seen in analogy with the first-order phase transition of an atomic or molecular Van der Waals gas.  In the nuclear case an important role for the interaction between nucleons is played by two-pion exchange processes which have qualitative features in common with the classical Van der Waals potential.}

\item{In isospin-asymmetric,  neutron-rich matter the coexistence region of the liquid-gas phase transition shrinks systematically with decreasing fractions of protons relative to neutrons.  The phase transition disappears as the system becomes unbound.  This is relevant for the matter that forms the bulk of neutron stars.}

\item{Advanced many-body calculations are capable of providing successful descriptions of the nuclear liquid-gas phase transition.  These include modern approaches based on chiral effective field theory,  the low-energy realisation of quantum chromodynamics which exhibits spontaneous and explicit breaking of chiral symmetry.  The functional renormalization group is a valuable method to treat fluctuations nonperturbatively in chiral nucleon-meson models.}
\end{itemize}
\begin{ack}[Acknowledgments]
~This work has been partially supported by the Excellence Cluster ORIGINS of the German Science Foundation (DFG).  Instructive comments by Peter Braun-Munzinger are gratefully acknowledged.
\end{ack}

\begin{ack}[References]

Andronic, A.  Braun-Munzinger, P., Redlich, K.,  and Stachel, J. , (2018). {\it Nature} {\bf 561},  321.\\
Aoki, S., and Doi, T. , (2023).  In: {\it Handbook of Nuclear Physics},Volume\,3, 1787.  Eds: Tanihata, I., Toki, H., and Kajino, T. , \\Springer Nature, Singapore.\\
Averbeck, R., Holzmann, R., Metag, V. , and Simon, R.S. , (2003).  {\it Physical Review} {\bf C67},  024903.\\
Beccatini, F., Cleymans, J., Ker\"anen, A., Suhonen, E., and Redlich, K. , (2001).  {\it Physical Review} {\bf C64},  024901.\\
Berges, J., Tetradis, N., and Wetterich, C. , (2002).  {\it Physics Reports} {\bf 363},  223.\\
Bernard, V., Kaiser, N., and Mei{\ss}ner, U.-G. , (1995).  {\it Int. Journal of Modern Physics} {\bf E4},  193.\\
Bertsch, G., and Siemens, P.J. , (1983).  {\it Physics Letters} {\bf B126},  9.\\
Boal, D.H. , (1986). {\it Nuclear Physics} {\bf A447},  479.\\
Bondorf, J.P., Donangelo, R., Mishustin, I.N., Pethick, C.J.,  Schulz, H. , and Sneppen, K. , (1985). {\it Nuclear Physics} {\bf A443},  321.\\
Bonnet, E., {\it et al.} , (2025). {\it Physical Review} {\bf C111}, 064616.\\
Borderie, B., {\it et al.} , (1999). {\it European Physics Journal} {\bf A6}, 197.\\
Borderie, B. , (2002). {\it Journal of Physics  G: Nucl. Part. Phys.} {\bf 28}, R217.\\
Borderie, B., and Rivet, M.F. , (2008). {\it Progress in Particle and Nuclear Physics} {\bf 61},  551.\\
Borderie, B., and Frankland, J.D. , (2019). {\it Progress in Particle and Nuclear Physics} {\bf 105},  82.\\
Borderie, B., {\it et al.} (2013). {\it Physics Letters} {\bf B723}, 140.\\
Braun-Munzinger, P., Rustamov, A., and Xu, N., (2026).  {\it arXiv:2601.18666 [nucl-ex]}.\\
Chomaz, P., and Gulminelli, F. , (2002).  In: {\it Dynamics and Thermodynamics of Systems with Long-Range Interactions}, 
{\it Lecture Notes in Physics (Springer)} {\bf 602}, 69-129.\\
Csernai, L.B.,  and Kapusta, J.L. , (1989).  {\it Physics Reports} {\bf 131},  223.\\
Drews, M., and Weise, W. , (2017).  {\it Progress in Particle and Nuclear Physics} {\bf 93},  69.\\
Drischler, C., Holt, J.W., and Wellenhofer, C. , (2021). 
{\it Ann. Review of Nuclear and Particle Sciences} {\bf 71}, 403.\\
Ducoin, C., Chomaz, P., and Gulminelli, F. , (2006).  {\it Nuclear Physics} {\bf A771},  68.\\
Durand, D., Suraud, B., and Tamain, B. , (2001).  {\it Nuclear Dynamics in the Nucleonic Regime},  Inst. of Phys. Publ.,  Bristol.\\
Elliot, J.B., Lake, P.T., Moretto, L.G., and Phair, L. , (2013). {\it Physical Review} {\bf C87}, 054622.\\
Epelbaum, E., Hammer, H.-W., and Mei{\ss}ner, U.-G. , (2009).  {\it Review of Modern Physics} {\bf 81}, 1773.\\
Eser, J., and Blaizot, J.-P. , (2024). {\it Physical Review} {\bf C109}, 045201.\\
Fiorilla, S., Kaiser, N., and Weise, W. , (2012).  {\it Nuclear Physics} {\bf A880},  65.\\
Floerchinger, S., and Wetterich, C. , (2012). {\it Nuclear Physics} {\bf A890-891},  11.\\
Friedman, B., and Pandharipande, V.R. , (1981). {\it Nuclear Physics} {\bf A361},  502.\\
Fritsch, S., Kaiser, N., and Weise, W. , (2002).  {\it Physics Letters} {\bf B545},  73.\\
Gell-Mann, M., Oakes, R., and Renner, B. , (1968). {\it Physical Review} {\bf 175},  2195.\\
Goldstone, J. , (1961).  {\it Nuovo Cimento} {\bf 19},  151.\\
Goodman, A.L., Kapusta, J.I., and Mekjian, A.Z. , (1984). {\it Physical Review} {\bf C30}, 851.\\
Gross, D.H.E. , (1990). {\it Reports on Progress in Physics} {\bf 53}, 605.\\
Gross, D.H.E. , (1993). {\it Progress in Particle and Nuclear Physics} {\bf 30},  155.\\
Hofmann, H. , (2008). {\it The Physics of Warm Nuclei},  Oxford University Press.\\
Holt, J.W., Rho, M., and Weise, W. , (2016). {\it  Physics Reports} {\bf 621},  2.\\
Jaqaman, H., Mekjian, A.Z., and Zamick, L. , (1983). {\it Physical Review} {\bf C27}, 2782.\\
Kadanoff, L.P. , {\it et al.} ,  (1967).  {\it Review of Modern Physics} {\bf 39}, 395.\\
Kadanoff, L.P. , (2000).  {\it Statistical Physics},  Ch. 11,  World Scientific, Singapore.\\
Kaiser, N., Gerstend\"orfer, S., and Weise, W. , (1998). {\it Nuclear Physics} {\bf A637},  395.\\
Kohn, W. , and Luttinger, J.M. , (1960) {\it Physical Review} {\bf 118}, 41.\\
Lamb, D.Q., Lattimer, J.M., Pethick, C.J., and Ravenhall, D.G. , (1978). {\it Physical Review Letters} {\bf 41}, 1623.\\
Landau, L.D. , and Lifshitz, E.M. , (1958).  {\it Statistical Physics}, Pergamon Press, Oxford.\\
Lee, B.W. , (1972).  {\it Chiral Dynamics},  Gordon and Breach Science Publishers, New York.\\
Luttinger, J.M.,  and Ward,  J.C. , (1960) {\it Physical Review} {\bf 118}, 1417.\\
Moretto, L.G., and Phair, L. , (2017).  {\it Journal of Physics: Conf.  Series} {\bf 863},  012055.\\
Nambu, Y., and Jona-Lasinio,G. , (1961). {\it Physical Review} {\bf 124}, 246.\\
Panagiotou, A.D., Curtain, W.M., Toki, H., Scott, D.K., and Siemens, P.J. , (1984).  
{\it Physical Review Letters} {\bf 52}, 496.\\
Pochodzalla, J., {\it et al.} , (1995).  {\it Physical Review Letters} {\bf 75}, 1040.\\
Rios, A. , (2010).  {\it Nuclear Physics} {\bf A845},  58.\\
Sauer, G., Chandra, H., and Mosel, U. ,  (1976). {\it Nuclear Physics} {\bf A264},  221.\\
Schwabl,  F. , (2006).  {\it Statistical Mechanics}, Ch. 5, Springer, Berlin \& Heidelberg.\\
Siemens, P.J. , (1983). {\it Nature} {\bf 305}, 410.\\
Weinberg, S. , (1990).  {\it Physics Letters} {\bf B251},  288.\\
Weinberg, S. , (1991).  {\it Nuclear Physics} {\bf B363},  3.\\
Wellenhofer, C., Holt, J.W., Kaiser, N., and Weise, W. , (2014). {\it Physical Review} {\bf C89}, 064009.\\
Wellenhofer, C., Holt, J.W.,  and Kaiser, N. , (2015). {\it Physical Review} {\bf C92}, 015801.\\
Wetterich, C. , (1993). {\it Physics Letters} {\bf B301},  90.\\ 

\end{ack}

\newpage

\end{document}